\documentclass[lettersize,journal]{IEEEtran}
\usepackage{amsmath,amsfonts,amssymb}
\usepackage{algorithm}
\usepackage{algpseudocode}
\usepackage{caption}
\usepackage{subcaption}
\usepackage{tabularray}
\usepackage{array}
\usepackage[caption=false,font=normalsize,labelfont=sf,textfont=sf]{subfig}
\usepackage{textcomp}
\usepackage{stfloats}
\usepackage{url}
\usepackage{verbatim}
\usepackage{graphicx}
\usepackage{cite}
\usepackage{svg}
\usepackage{color}
\begin{document}

\title{Distributed Entanglement Distribution Using Multiple Entanglement Sources in WDM-based Quantum Optical Networks}


\author{Anuj Agrawal$^{\dag,*}$, Abhay Singh Dulta$^{\dag}$, Bhaskar Kanseri$^{\S}$
\thanks{This work is supported by the National Quantum Mission (NQM), Ministry of Science and Technology, India, and the Indian Institute of Technology (IIT) Delhi, India.}
\thanks{$^{\dag}$A. Agrawal and A.S. Dulta are with the Experimental Quantum Interferometry and Polarization (EQUIP) Group, Department of Physics, IIT Delhi, New Delhi, India (e-mail: anujagrawal@ieee.org$^*$, phz228454@iitd.ac.in).}
\thanks{$^{\S}$B. Kanseri is with the Experimental Quantum Interferometry and Polarization (EQUIP) Group, Department of Physics, and Optics and Photonics Centre (OPC), IIT Delhi, New Delhi, India (e-mail: bkanseri@physics.iitd.ac.in).}
\vspace{-2em}}



\maketitle

\begin{abstract}

Quantum network implementations using single spontaneous parametric downconversion (SPDC)-based broadband entangled photon pair source (EPPS) have been reported recently. Here, leveraging the wavelength-correlation between entangled photon pairs, the traditional wavelength division multiplexing (WDM) method is utilized to route photons based on their wavelengths. From single EPPS, entangled photon pairs are distributed in a centralized way to different node pairs in the network. However, the number of nodes pairs that can be entangled in a network is limited by the number of entangled wavelength pairs that an EPPS can generate. To entangle a higher number of node pairs in a network, multiple EPPSs can be employed. In this work, we present a WDM-based entanglement distribution approach using multiple EPPSs in multi-hop repeaterless mesh optical networks. We experimentally characterize two EPPSs developed in-house and consider multiple such EPPSs in the network to perform network-level simulations. We consider heterogeneous entanglement demands requiring different entanglement bit (ebit) rates and entanglement visibility. For each entanglement demand, EPPS placement/selection, wavelength-pair assignment, and photon pair routing are performed considering the degradation in both ebit rate and visibility with fiber length and hops in the network. Two main findings of this study include: (i) a hybrid approach of one-photon (OP) and both-photon (BP) entanglement distribution provides higher flexibility in multi-EPPS placement and (ii) distributed entanglement distribution using multiple EPPSs enables better entanglement resource utilization and higher entanglement demand acceptance as compared to centralized entanglement distribution.

\end{abstract}
\vspace{-0.3em}
\section{Introduction}

Entanglement is the quantum resource that is continuously consumed to run several quantum network applications including distributed quantum computing, clustered quantum computing, quantum-enhanced navigation, very-long-baseline interferometry, quantum key distribution, among others \cite{qnar}. Thus, distribution of this quantum resource, i.e., entanglement distribution at required rate and quality, is the fundamental task in quantum networks. Among the envisioned quantum network applications \cite{qnar,Kanseri:26}, most of the applications can be realized using Bell-state entanglement (also referred as Bell pair) \cite{bell1964einstein}. Spontaneous parametric downconversion (SPDC)-based polarization-entanglement source has been commonly employed as a Bell pair source or entangled photon pair source (EPPS) in quantum networking studies \cite{Wengerowsky2018,wang2022dynamic} due to its ease of generation, manipulation, and compatibility with off-the-shelf optical devices.

\begin{figure}[t]
	\centering
	\includegraphics[width=0.9\linewidth]{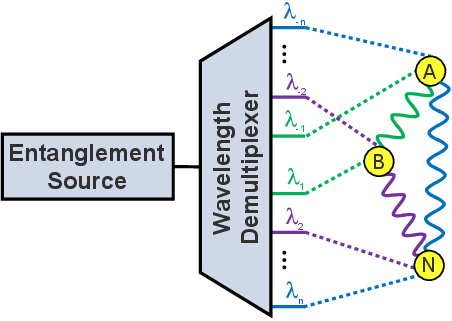}
	\caption{An example of WDM-based entanglement distribution using single broadband EPPS \cite{Wengerowsky2018}. Here, the entanglement source spectrum is first demultiplexed into different wavelengths and then entangled photon pairs at correlated wavelength-pairs (shown in same color) are routed towards the two entangling nodes (connected by a wiggly line of same color as the wavelength-pair color).}
	\label{fig:entdist}
	\vspace{-2.1em}
\end{figure}

Wavelength division multiplexing (WDM) has been of great utility in the existing classical optical networks for several decades. WDM is now being leveraged in quantum optical networks too, where a multi-node repeaterless quantum network can be established using single broadband EPPS by transmitting entangled photons at different wavelengths to different nodes \cite{Wengerowsky2018,wang2022dynamic}. Fig. \ref{fig:entdist} shows an example of WDM-based entanglement distribution using single broadband EPPS, where the output of an entanglement source is first demultiplexed into different wavelengths (also referred to as `channels') and then photons at correlated wavelengths are routed towards different nodes in the network. As per the law of conservation of energy, in the SPDC process, the entangled photons are correlated in wavelengths such that a photon at wavelength $\lambda_{c+\delta c}$ has its corresponding paired photon at $\lambda_{c-\delta c}$, where $\lambda_{c}$ is the central wavelength of the SPDC output spectrum. Thus, one node pair in the network can be entangled utilizing one correlated wavelength pair. For example, to entangle node A and node B, photons at $\lambda_{-1}$ and $\lambda_{1}$ (shown in green color) can be routed to node A and node B, respectively, as shown in Fig. \ref{fig:entdist}. Similarly, the wavelength pairs $\lambda_{2}-\lambda_{-2}$ and $\lambda_{n}-\lambda_{-n}$ can be used to entangle node pairs B$-$N and A$-$N, respectively. These node pairs are referred to as the `entangling nodes' or `entangling node pair'. Depending on the quantum network application, the distributed entangled photons can then either be measured immediately (such as to perform quantum key distribution) or be stored in quantum memories (such as to perform clustered quantum computing). In addition, depending on the quantum network application, different entangling node pairs require different entanglement bit (ebit) rates and entanglement quality \cite{qnar} (measured in terms of visibility, fidelity, negativity, S-parameter, etc.). In this work, we use visibility as the measure of entanglement quality and consider different visibility thresholds for different entanglement demands as our experimental EPPS characterization (explained in Section III) is in terms of visibility.

For WDM-based entanglement distribution between all the $n\choose 2$ unique node pairs in a network of $n$ nodes using one EPPS, $n\choose 2$ wavelength pairs need to be generated by the EPPS (see Fig. \ref{fig:entdist}). Thus, the number of node pairs that can be entangled in a network using one EPPS is limited by the number of wavelength pairs generated by the EPPS. For a fixed spectral granularity (e.g., 100 GHz), this means that with an increase in the number of entangling nodes in a network, the EPPS output spectrum should be increasingly broader. While EPPS bandwidth can be varied to an extent with device-level reconfiguration by temperature and wavelength tuning in the type-0 SPDC process \cite{steinlechner2014efficient}, it affects the optimality of EPPS due to simultaneous undesired variations in central wavelength, photon count rates, among others. Another approach to increase the number of wavelength pairs is by using finer spectral granularity de/multiplexing (e.g., 50 GHz, 25 GHz), however, it reduces the photon counts per wavelength and hence the resulting ebit rate for each node pair. Moreover, this centralized approach of entanglement distribution, where single broadband EPPS distributes entangled photon pairs to all the node pairs (some of which could be located far away from the EPPS location) in the network would result in a significantly high loss of entangled photon pairs enroute the links and nodes in the network.
To circumvent these limitations, multiple narrower-band EPPSs can be deployed distributively to entangle the same number of node pairs in the network, which is the focus of this work. In WDM-based quantum networks, photons are routed solely on the basis of their wavelengths \cite{Wengerowsky2018}. With only one EPPS in a network, no challenge is encountered at nodes (consisting of wavelength de/multiplexers and/or wavelength-selective switches (WSSs)) to make wavelength-based routing decisions. However, with multiple EPPSs sharing fiber links in a common network, to avoid wavelength contention, the output spectra of different EPPSs must not overlap with each other (further explained in Section IV and depicted in Fig. \ref{fig:multisource_spectra}). With multiple EPPSs to be operated simultaneously in a network, where should different EPPSs be placed in the network for optimal performance is an important question to be addressed. Further, selecting one suitable EPPS among others to provision each entanglement demand is another crucial problem as it can impact the overall entanglement distribution rate and visibility in the quantum network.

In this work, we present a distributed entanglement distribution approach using multiple SPDC-based EPPSs in WDM-based quantum optical networks. We first experimentally characterize entangled photon pair generation rates and entanglement visibilities using two type-0 SPDC-based EPPSs developed in-house (explained in Section III) and then perform network-level simulations considering multiple such EPPSs in the quantum optical network. From the demultiplexed output spectrum of the EPPS (shown in Fig. \ref{fig:singlespectra}), it can be observed that different wavelength pairs offer different ebit rates and/or visibilities (explained further in Section III). Thus, we perform entanglement demand-aware wavelength pair assignment to provision heterogeneous entanglement demands representing different quantum applications for which entanglement is distributed in the same network. While initial repeaterless and trusted-node-free WDM-based entanglement distribution demonstrations have been based on star network topology \cite{Wengerowsky2018, joshi2020trusted, wang2022dynamic}, mesh networks are preferred and are widely deployed due to its merits such as path diversity, better fault-tolerance, flexible routing, better resource utilization, among others. However, WDM-based entangled photon routing over multi-hops in multi-node mesh networks becomes more complex than star networks, and the mesh network nodes should have wavelength-based add/drop/bypass capability. In this work, we consider mesh optical networks of varying connectivity with multiple EPPSs placed at different nodes in the network to analyze their performance.

A node having an EPPS placed at it is referred to as a `source node' in this paper. Each node pair in the network can be entangled by either using an EPPS available at one of the two entangling nodes or using an EPPS available at some other node in the network. The former approach simplifies photon routing since in this case, one photon from each entangled photon pair can be routed within the source node and only the other photon needs to be routed over the network towards the second entangling node. This is advantageous in 2-node (point-to-point) studies, however, to entangle all the node pairs in a network, this approach requires an EPPS to be available at all nodes except one node in a multi-node mesh network. On the other hand, using an EPPS placed at none of the two entangling nodes (which is the case in star networks) would require routing both the entangled photons of each pair from EPPS over the network, in which case routing becomes more complex and the resulting photon pair loss would be multiplicative (or additive in decibel scale) over the two routes that each photon of the pair traverse. We study both these photon distribution approaches (referred to as `one photon' (OP) and `both photons' (BP) distribution schemes) and compare several scenarios of EPPS placements for WDM-based entanglement distribution in repeaterless multi-hop mesh quantum optical networks. For a comprehensive analysis, we consider several network topologies such as ring, full-mesh, and random partial-mesh network topolgies obtained using Erdős--Rényi graph model \cite{erdos1959random}. For EPPS placement, we consider the graph-theoretical concepts of `betweenness centrality of a vertex' and `dominant set'. Entangled photon pair generation rates (coincidence counts) are obtained experimentally using two EPPSs developed in-house and then demand-, loss-, and visibility-aware entanglement distribution is performed considering multiple such EPPSs in the network.

Main contributions of this paper are summarized as: (a) A distributed entanglement distribution approach using multiple EPPSs in WDM-based quantum optical networks is proposed; (b) A candidate node architecture for WDM-based entanglement distribution in mesh optical networks using multiple EPPSs is presented; (c) Estimated reduction in both entanglement rate and visibility with distance and hops is considered while provisioning heterogeneous entanglement demands; and (d)  A comprehensive analysis considering different photon routing approaches and centralized/distributed EPPS placements in several types of network topologies is done.

\section{Related Work}

This paper is focused on repeaterless and trusted-node-free entangled photon pair (i.e., bipartite entanglement) distribution using a passive WDM approach, hence we discuss the related works which this study builds up on accordingly in this section. Beyond 2-node point-to-point studies, the first fully passive WDM-based bipartite entanglement distribution using a single EPPS was demonstrated in \cite{Wengerowsky2018} for a 4-node star network topology, where a separate node (referred to as `network provider' in \cite{Wengerowsky2018}) distributes entangled photon pairs at different wavelengths to different node pairs, as represented in Fig. \ref{fig:entdist}. The EPPS used here \cite{Wengerowsky2018} is a Sagnac-loop configured type-0 SPDC-based source that generates six wavelength pairs of 100 GHz granularity, which entangle all the $4\choose2$$=6$ node pairs in the 4-node star network. In \cite{wang2022dynamic}, a similar setup is used, however, using a configurable optical fiber switch (OFS) between the demultiplexing and multiplexing stages, it enables dynamicity to reconfigure the network. Here, a similar EPPS that generates 15 wavelength pairs of 100 GHz granularity has been used that can entangle up to $6\choose2$$=15$ node pairs in the considered 6-node star network. While these works laid a foundation for WDM-based entanglement distribution in quantum optical networks, the transmission of photons was still point-to-point (from a central node to other nodes) in a hub and spoke model. Moreover, using a separate node as `network provider' (i.e., a separate location where an EPPS is installed) may not be as pragmatic approach as co-locating the EPPS at one of the network nodes (i.e., entangling nodes) themselves. In case of multiple EPPSs placed at different nodes in mesh networks, a node can use an EPPS available locally to entangle itself with another node in the network in which case only one photon of the pair needs to be routed over the network. However, in multi-hop mesh networks, photons need to be bypassed through intermediate nodes. In \cite{jsac, mcaleese2025strategies}, WDM-based entanglement distribution in mesh optical network using single, however, different types of EPPSs based on dual-SPDC and quantum discord, are presented, where an EPPS placed at one of the entangling nodes has been considered.

Using single EPPS, the number of nodes that can be entangled is limited by the number of wavelength pairs that an EPPS generates, as explained in Section I, unless further de/multiplexing based on other degrees of freedom such as polarization \cite{joshi2020trusted} or finer spectral granularities (e.g., 50/25/12.5 GHz) are used, which reduce the ebit rates per wavelength pair or some wavelength-pairs may remain unused if the node pairs to be entangled are located farther from the EPPS location. We argue in this paper that the centralized approach of entanglement distribution from a single broadband EPPS causes a huge loss of the finite quantum entanglement resource and hence we propose a distributed approach of entanglement distribution using multiple narrower-band EPPSs, where several EPPSs are placed distributively at different nodes in the network. One of the EPPSs and wavelength pairs are then selectively used to entangle each node pair in the network, as per the entanglement demands. To the best of our knowledge, this is the first work that addresses WDM-based entanglement distribution in quantum optical networks using multiple EPPSs.

In entanglement distribution studies that consider quantum-memory based quantum repeaters, the entanglement quality not only degrades with distance but also with the operations such as entanglement swapping (used for long distance) and entanglement purification (used to improve entanglement fidelity). In addition, quantum transduction \cite{caleffi2025quantum} and its efficiency are other important factors that impact the ebit rate and quality of entanglement. Thus, entanglement distribution analysis should be done considering the effect of such operations that affect the resulting ebit rate and fidelity. In \cite{pant2019routing}, quantum repeater protocols are studied and it has been found that routing entanglement through multiple paths in a quantum network improves entanglement distribution rate. However, entanglement fidelity is rather considered as a constraint. In \cite{li2022fidelity}, it is highlighted that fidelity is rarely considered in the entanglement routing studies and hence the authors proposed a method for entanglement routing with guaranteed-fidelity. First-generation quantum repeaters \cite{muralidharan2016optimal}, deterministic entanglement generation and a constant capacity for each quantum channel are considered in \cite{li2022fidelity}. In \cite{gu2024fendi}, the importance of considering both the ebit rate and entanglement fidelity for entanglement distribution is further emphasized, where the authors study the rate-fidelity trade-off considering first-generation quantum repeater networks. A survey on several aspects of entanglement routing in quantum repeater equipped quantum networks is done in \cite{abane2025entanglement}. A comprehensive survey underscoring the protocol-level fundamental aspects in quantum repeater equipped quantum networks is presented in \cite{illiano2022quantum}. Here, the classical-inspired methods, protocols, and layering approaches proposed for entanglement-based quantum networks are reviewed and their suitability for the quantum internet is flagged. As our work is focused on the approach of repeaterless WDM-based broadband entanglement distribution \cite{Wengerowsky2018} from probabilistic type-0 SPDC-based EPPSs, we defer the readers interested in entanglement distribution in quantum repeater networks to \cite{pant2019routing, li2022fidelity, gu2024fendi, abane2025entanglement, illiano2022quantum}. In our work, the network nodes are not equipped with quantum repeaters. Here, the intermediate nodes in mesh optical network passively route the photons based on their wavelength towards the destination node, as explained in section IV.

In classical WDM-based optical networks, any of the available wavelengths can be assigned for lightpath establishment \cite{zang2000review}. However, for WDM-based entanglement distribution in quantum optical networks, wavelength pair assignment should be performed based on each entanglement demand since each wavelength pair offers different entanglement rates (characterized by coincidence counts) and entanglement visibilites. In \cite{trapateau2015multi}, it has been experimentally demonstrated that entanglement visibilities may vary for different wavelength pairs as well as with different demultiplexing technologies. Thus, for wavelength-pair assignment in quantum optical networks, it must be ensured for each entanglement demand that the required ebit rate and entanglement visibility are achievable on reaching the entangling node pair. In this work, we consider both the entangled photon pair rates and the visibilities for different wavelength pairs and their estimated reduction over fiber distances and hops while provisioning the entanglement demands in multi-hop mesh networks, as explained in Section IV. However, unlike first-generation quantum repeater based networks where entanglement quality is significantly impacted by swapping and purification operations, we study the impact on entanglement visibility degradation due to the increasing accidental-count contribution with increase in fiber length, as explained in the Section IV.


\section{Characterization of Entangled Photon Pair Source (EPPS)}

\begin{figure}[h]
	\centering
	\includegraphics[width=0.9\linewidth]{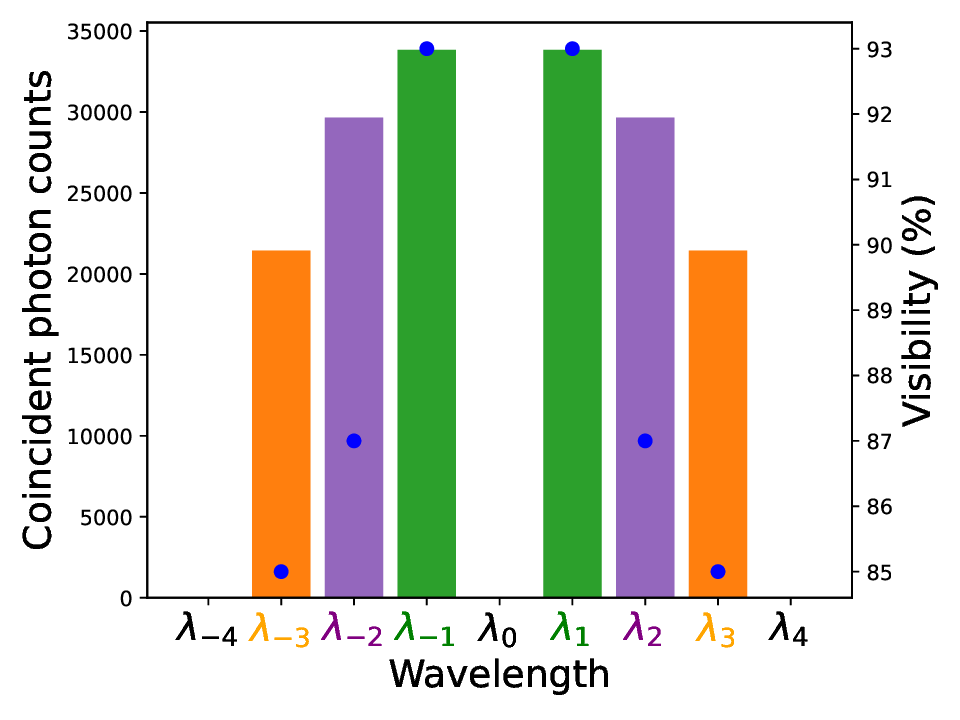}
    \vspace{-0.5em}
	\caption{Experimental characteristics of the three wavelength pair EPPS (3WPS) considered in this work. Here, coincident photon counts for different wavelength pairs (shown in same color) are represented by bars and their entanglement visibilities are represented by blue dots. The central wavelength ($\lambda_0$) is the point of degeneracy.}
	\label{fig:singlespectra}
	\vspace{-1em}
\end{figure}

We experimentally obtain characteristics of two EPPSs developed in-house \cite{kanseri2026patent}. The developed EPPSs are Sagnac-loop configured type-0 SPDC-based (periodically poled lithium niobate waveguide) sources similar to that used in \cite{Wengerowsky2018} and \cite{wang2022dynamic}. The developed EPPSs are optimized for coincidence photon counts. The first EPPS generates three wavelength pairs (as shown in Fig. \ref{fig:singlespectra}) and the second EPPS generates five wavelength pairs.  These coincidence counts for different wavelength pairs are considered as the entanglement generation rates at source. As highlighted in \cite{trapateau2015multi}, our experiment confirms that different entanglement visibilities (shown by blue circles in Fig. \ref{fig:singlespectra}) may be observed for different wavelength pairs. Thus, while allocating a wavelength pair, a demand-aware entanglement distribution approach should select a wavelength pair based on its visibility at source and the estimated visibility degradation en-route.

It should be noted that the central wavelength ($\lambda_{0}$ in Fig. \ref{fig:singlespectra}), referred to as the `point of degeneracy' is not used \cite{Wengerowsky2018,wang2022dynamic} since the entangled photons at the central wavelength are indistinguishable in all the degrees of freedom including wavelength and hence cannot be separated deterministically. The wavelength pairs beyond $\lambda_{-3}-\lambda_{3}$ in Fig. \ref{fig:singlespectra} have further lower coincidence photon counts and entanglement visibilities below the threshold ($71\%$). Thus, wavelength pairs beyond $\lambda_{-3}-\lambda_{3}$ are filtered out from this three wavelength pair EPPS (3WPS) spectrum. Since the coincidence counts  and visibility measurements shown in Fig. \ref{fig:singlespectra} already include the effect of inefficiencies (dark count rate, detection efficiency, etc.) of the superconducting nanowire single-photon detector (SNSPD) used in our experiment, its effect is not incorporated again while provisioning entanglement demands. Our second EPPS provides five wavelength pairs (referred to as five wavelength pair EPPS (5WPS) in this work), higher coincidence counts per wavelength pair as compared to 3WPS, and almost constant visibility ($\sim 92\%$). Though the visibility observed for 5WPS is the same for all the wavelength pairs, the node pairs to which they are assigned define the effective visibility based on the distance/s from the selected EPPSs to the entangling node pairs, as explained next in Section IV. The characteristics of the 5WPS are given in Table \ref{tab3}. We consider both the 3WPSs and 5WPSs for network-level performance evaluation in Section V.

\section{Distributed Entanglement Distribution with Multiple EPPSs}

In classical optical networks, light signals at same wavelength cannot co-propagate over same fiber without interfering. Moreover, unidirectional amplification using erbium doped fiber amplifiers (EDFAs) in long-distance optical networks prohibit bidirectional communication on a single optical fiber. Thus, a pair of fiber is utilized for bidirectional communication in classical optical networks. However, in quantum optical networks, entangled photons of same wavelength can travel together in the same or opposite direction as their interference is possible only when they are indistinguishable (in all the degrees of freedom) and that too using a special interferometric setup (such as Bell state measurement (BSM)). Moreover, classical amplification is absent (since impossible) in quantum optical networks. Thus, bidirectional communication of entangled photons generated by different SPDC-based EPPSs is possible over single optical fiber in quantum optical networks. However, for WDM-based entanglement distribution, as represented in Fig. \ref{fig:entdist}, entangled photons are routed towards different nodes in the network merely on the basis of their wavelength. Thus, even though same-wavelength photons can traverse the same optical fiber without interference, it is not possible for WSSs or wavelength de/multiplexers to distinguish among multiple photons of same wavelength generated by different EPPSs. Thus, the spectrum of different EPPSs used in the network for WDM-based entanglement distribution must not overlap with each other.

\begin{figure}[h]
	\centering
	\includegraphics[width=0.99\linewidth]{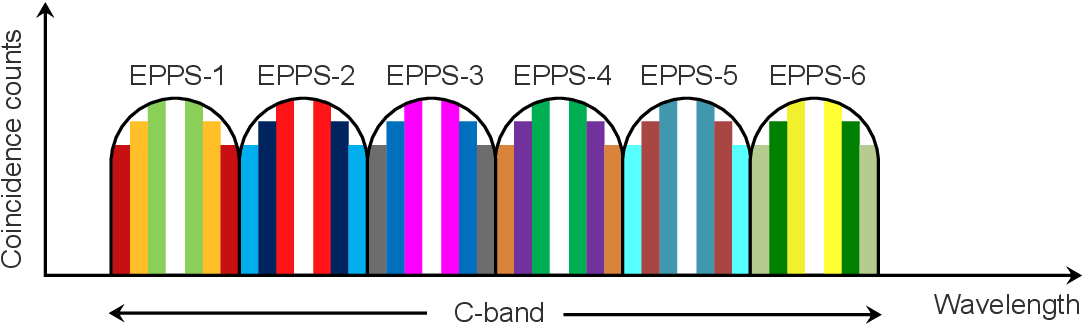}
	\caption{Wavelength-multiplexing of six 3WPSs in optical C-band (considering $6\times 7 = 42$ wavelengths of $100$ GHz granularity occupying $4.2$ THz out of total $\sim 4.4$ THz C-band). Same-color bars indicate wavelength correlation between entangled photon pairs.}
	\label{fig:multisource_spectra}
	\vspace{-0.9em}
\end{figure}

In Fig. \ref{fig:multisource_spectra}, non-overlapping wavelength-multiplexed spectra of six 3WPSs (as per Fig. \ref{fig:singlespectra}) is shown. Here, coincidence counts bars of same color indicate the correlated wavelengths pairs at which entangled photon pairs are generated. Note that the central wavelength of the output spectrum of SPDC-based EPPS can be tuned by tuning the pump wavelength and/or temperature of the crystal/waveguide during the SPDC process while achieving almost similar coincidence counts \cite{steinlechner2014efficient}. Thus, simultaneous operation of six 3WPSs, as shown in Fig. \ref{fig:multisource_spectra}, is feasible without any routing ambiguity for WSSs or wavelength de/multiplexers used for WDM-based entanglement distribution in quantum optical networks. A total of 18 wavelength pairs (represented by same-color bars) obtained using six 3WPSs, as shown in Fig. \ref{fig:multisource_spectra}, can be used to entangle up to 18 node pairs in a quantum network. In this work, we consider transmission in only the conventional (C)-band (1530-1565 nm) of $\sim 4.4$ THz bandwidth, where maximum six 3WPSs (each occupying 7 wavelengths of 100 GHz granularity) can operate simultaneously without any spectral overlapping. Similarly, maximum four 5WPSs (each occupying 11 wavelengths) can operate simultaneously without any spectral overlapping within C-band. However, more such EPPSs can be used in case of multi-band optical networks and the presented work remains applicable, without loss of generality, in that case.

\begin{figure}[t]
	\centering
	\includegraphics[width=0.99\linewidth]{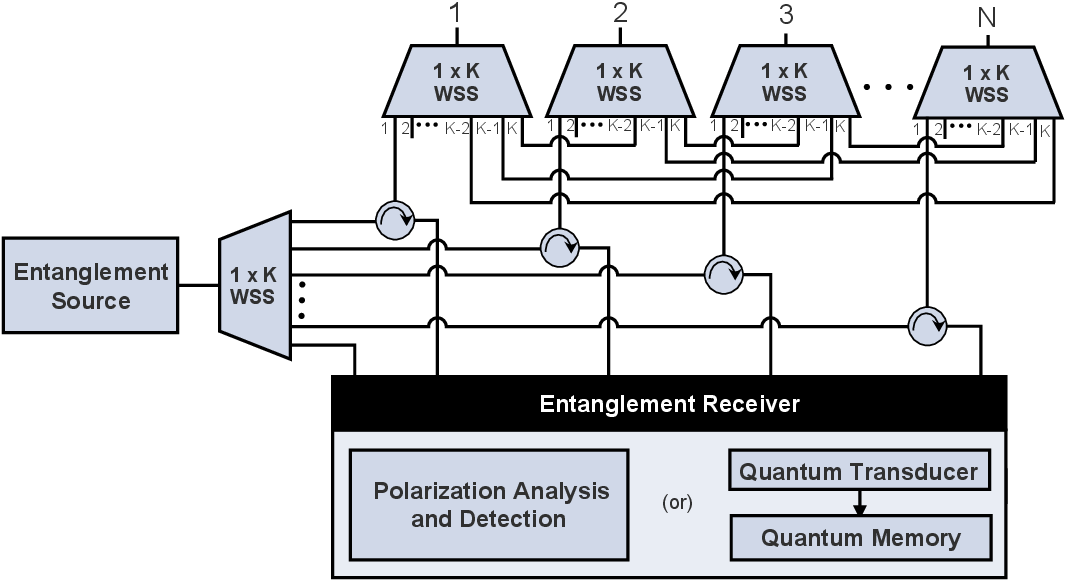}
	\caption{Node architecture for WDM-based entanglement distribution with multiple EPPSs in multi-hop mesh quantum optical networks. Optical circulators are represented by circles with circulation direction denoted by arrows inside them.}
	\label{fig:nodearch}
	\vspace{-0.9em}
\end{figure}

\begin{figure}[h]
	\centering
	\includegraphics[width=0.7\linewidth]{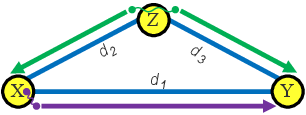}
	\caption{Schemes of photon pair distribution: when source node (X) is one of the entangling nodes (X-Y), only one photon (OP) is transmitted over the network, as represented in purple color; and when source node (Z) is none of the entangling nodes (X-Y), both photons (BP) are transmitted over the network, as represented in green color.}
	\label{fig:photonpair}
	\vspace{-0.9em}
\end{figure}

With multiple EPPSs in a WDM-based multi-node mesh optical network, a node can either transmit, receive, or bypass photons based on their wavelength. To this end, we present a candidate node architecture in Fig. \ref{fig:nodearch}. Here, the EPPS output is first demultiplexed using a WSS (which enables electronically re/configurable multi-wavelength selection on any desired port) and then different wavelengths are routed towards different nodes in the network (as per the photon routing and wavelength-pair assignment discussed later in this section). Depending on the location of the EPPSs, an EPPS can be used to entangle a node at which it is placed with another node in the network, i.e., source node is one of the entangling nodes. In this case, only one photon needs to be transmitted out of the source node over the network towards the other entangling node, and the second paired photon is sent to the entanglement receiver within the source node. Notice that one of the output ports of the WSS (in left) is connected to the entanglement receiver within that node to enable this case. This is referred to as `one-photon' (OP) distribution scheme in this work and is depicted in Fig. \ref{fig:photonpair} in purple color. The design of entanglement receiver is application-specific and could consist of either a polarization analysis and detection module or a quantum transducer and quantum memory module. We restrict this work to entanglement distribution and coincidence detection to calculate the distributed ebit rates and visibilities after traversing the selected routes. However, application-specific effective rates such as for memory-based applications, the efficiency of quantum transduction \cite {caleffi2025quantum}, the rate of entangled qubit pair storing and retrieving in/from quantum memories, memory decoherence, memory buffer size, classical communication latency, etc. \cite{vasan2025control}, would further scale down the usable rates as per different hardware-level inefficiencies (which we consider to be out of the scope of this work).

When an EPPS is used to entangle nodes other than the node at which it is placed, both the photons are routed out of the source node. Thus, source node is none of the two entangling nodes in this case. This is depicted in Fig. \ref{fig:photonpair} in green color and is referred to as `both photon' (BP) distribution scheme in this work. The array of N interconnected WSSs (where, N is equal to the degree of the node at which an EPPS is placed) shown at top in Fig. \ref{fig:nodearch}, is used to enable photon routing using OP or BP through the desired fiber-link (as per the selected route) as well as to enable bypass of photons at intermediate nodes. For instance, a photon entering a WSS from fiber-link `1' can exit through its `K-1' port to bypass an intermediate node and traverse fiber-link `3' further. The optical circulators (OCs) enable bidirectional communication of photons through nodes in the network such that any incoming photon at a node from an EPPS placed at another node in the network is either bypassed or routed towards its entanglement receiver (depending on photon's destination) but never towards the EPPS shown at the left end. For those nodes in the network where no EPPSs are deployed, the node architecture consists of only the WSS-array directly (i.e., without circulators) connected with the entanglement receiver/s.

The entanglement visibility ($V$) for different wavelength pairs is calculated as 

\begin{equation}
    V = \frac{N_{c,max} - N_{c,min}}{N_{c,max} + N_{c,min}},
\end{equation}
where, $N_{c,max}$ and $N_{c,min}$ are the measured maximum and minimum coincidence counts, respectively, for a wavelength pair. Eq. (1) is applicable when calculating $V$ at source, i.e., connecting EPPS directly (with negligible channel length) with coincidence detection module and assuming no detector inefficiencies. However, as the entangled photons are distributed to different nodes in the network, due to fiber attenuation, either or both of the entangled photons from each pair may be lost en-route. The resulting photon-pair loss is calculated as the sum (product) of losses in dB (linear scale) accrued over the two different paths that each photon traverses in the network. Let $\eta_{L_1}$  and $\eta_{L_2}$ be the transmission efficiencies of the two paths that the photons travel due to fiber attenuation, given by $10^{-\alpha_{L_1}/10}$ and $10^{-\alpha_{L_2}/10}$, respectively. Here, $L_1$ and $L_2$ are the link distances that the first and the second photon traverse, respectively, and $\alpha$ is the fiber attenuation coefficient. Let $N_{acc}$ be the number of accidental coincidence counts due to detector dark counts and ambient light. Since coincidence detection requires both the photons to reach the two entangling nodes within a certain time interval, usually termed as coincidence window, the signal scales as $\eta_1\eta_2$. Therefore, the maximum and minimum coincidence counts become

\begin{equation}
N_{c,max}=\eta_1\eta_2N_{c,max}+N_{acc},
\end{equation}

and
\begin{equation}
N_{c,min}=\eta_1\eta_2N_{c,min}+N_{acc}
\end{equation}
On substituting $N_{c,max}$ and $N_{c,min}$ from Eq. (2) and Eq. (3) to Eq. (1) and simplifying, the visibility ($V(d)$) as a function of distance at the end nodes can be obtained as

\begin{equation}
V(d)=
\frac{
N_{c,max}-N_{c,min}
}{
N_{c,max}+N_{c,min}+\dfrac{2N_{acc}}{\eta_{L_1}\eta_{L_2}}
}
\end{equation}

Eq. (4) shows that as the fiber length increases, $\eta_{L_1}\eta_{L_2}$ decreases, causing the accidental-count ($N_{acc}$) contribution in the denominator to become more significant, thereby reducing the entanglement visibility. The losses due to OCs and WSSs at the two entangling nodes and the intermediate nodes (see Fig. \ref{fig:nodearch}) en-route traversed by photons further reduce the entanglement distribution rate and visibility. Let $\eta_{sw}$ be the switching efficiency given by $10^{-L_{sw}/10}$, where $L_{sw}$ is the total switching loss (due to OCs and WSSs) along the two routes. Here, it is assumed that the effect of a switching loss of $L_{sw}$ dB on $V(d)$ is equivalent to the effect of a fiber of length $L_{sw}/\alpha$ km. Incorporating the switching losses, Eq. (4) becomes

\begin{equation}
    V(d) = \frac{N_{c,max} - N_{c,min}}{N_{c,max} + N_{c,min} + \dfrac{2N_{acc}}{\eta_{L_1} \eta_{L_2}\eta_{sw}}}
\end{equation}
In case of BP distribution represented in green color in Fig. \ref{fig:photonpair}, where source node (Z) is none of the two entangling nodes (X-Y), $L_{1}=d_2$ and $L_{2}=d_3$. In case of OP distribution represented in purple color in Fig. \ref{fig:photonpair}, where source node (X) is one of the two entangling nodes (X-Y), $L_{1}=0$ (as this photon is sent to the entanglement receiver within node X) and $L_{2}=d_1$ as the other paired photon is sent across the network over a fiber of length $d_{1}$. In this work, we perform both loss- and visibility-aware provisioning of entanglement demands, as discussed next.

\begin{table}[!t]
\centering
\caption{Notations}
\vspace{-0.7em}
\label{tab:notations}
\begin{tabular}{p{0.3cm}p{7.1cm}} 
\hline
\hline
\multicolumn{1}{l}{Symbol} & \multicolumn{1}{l}{Description} \\ \hline
$\mathbb{N}$ & set of natural numbers \\
$\mathbb{P}$ & probability \\
$|\cdot|$ & Cardinality\\
$N$ & set of nodes, index $n$, $N\subset \mathbb{N}$\\
$E$ & set of links, index $e$, $E\subset \mathbb{N}$\\
$d_e$ & length of link $e$, $\forall e \in E$\\
$W$ & set of wavelengths available on each fiber link, index $w$, $W\subset \mathbb{N}$\\
$U_w$ & Binary variable, `$U_w=0$' if wavelength $w$ is available, `$1$' otherwise\\
$S$ & set of EPP sources, index $s$, $S\subset \mathbb{N}$\\
$\Lambda$ & set of central wavelengths of EPP sources' output spectra, index $s$, $\Lambda\subset W$\\
$x_s$ & location of EPP source $s$, $\forall s \in N$\\
$g$ & number of wavelengths per EPP source\\
$c_{gs}$ & photon counts at wavelength $g$ in the output spectra of an EPP source $s$, $ g \in W$\\
$Z_s$ & Binary variable, `$Z_s=1$' if all the wavelengths of the EPP source placed at $x_s$ are unavailable, `$0$' otherwise\\
$Q$ & set of entanglement demands, index $q$, $Q\subset \mathbb{N}$\\
$V$ & entanglement visibility\\
$R$ & ebit rate\\
$\Theta$ & a very large number\\
\hline
\hline
\end{tabular}
\end{table}

Quantum networks are envisioned to serve multiple quantum applications over the common network infrastructure. Different quantum applications have different requirements \cite{qnar}. From the entanglement distribution perspective, rate and quality (measured in terms of visibility in this work) of distributed entangled photon pairs are of predominant importance. As the entangled photon pairs obtained using an EPPS are finite and subject to losses, it is beneficial from an operator's perspective to allocate this quantum resource efficiently as per the entanglement demands. Let entanglement demands be denoted by $Q\{a_{q},b_{q},V_{q},R_{q}\}$, where $a_{q}$ and $b_{q}$ are the two entangling nodes demanding entanglement at an ebit rate of $R_{q}$ pairs per second with minimum visibility $V_{q}$. Thus, to distribute entangled photon pairs to different demands as per their requirements in a quantum optical network with multiple EPPSs, following tasks need to be done: (i) placement of EPPSs at node/s in the network, (ii) selection of one of the multiple EPPSs available in the network to serve each entanglement demand, (iii) one of the OP or BP distribution schemes is used depending on the available and selected EPPS, and (iv) assignment of one of the available wavelength pairs generated by the selected EPPS. These tasks aren't necessarily sequential but each can impact the overall network performance. To analyze the effect of EPPS placement and OP/BP distribution scheme, we first consider gradual placement of EPPSs in a given network as the entanglement demands arrive in the network. We consider an incremental traffic model \cite{zang2000review}, where entanglement demands ($Q$) of infinite holding times arrive randomly (uniform distribution) in the network. Table \ref{tab:notations} lists the notations used in this paper.

\setlength{\textfloatsep}{0.1cm}
\setlength{\floatsep}{0.1cm}
\begin{algorithm}[!t]
	\footnotesize
	\caption*{Algorithm 1: One Photon (OP) Distribution}
	\label{algo:flop}
	\textbf{Input:} $N, E, d_e, W, S, c_{gs}, \Lambda, g,$ $Q\{a_{q},b_{q},V_{q},R_{q}\}$, $s_{u}\leftarrow 0$, $U_w$, $Z_s$, $Q_{acc}\leftarrow 0$,  $y\gets 0$,  $x_s\gets 0\text{~} \forall s$\\
	\textbf{Output:} $Q_{acc}, s_{u}, x_{s}$
	\begin{algorithmic}[1]
		\For{$q\leftarrow$ $1$ to $|Q|$}
        \State $l_q \gets \Call{MinLossRoute}{a_q,b_q,N,E,d_e}$;
        \If {$(a_{q} \in x_s)$ $\land$ $(Z_{a_q}=0)$};
        \State $U_w,Z_{a_q},blk, R'_q \gets \Call{FFWavPair}{g,\Lambda_{a_q},c_{gs},l_{q},R_{q},V_{q},U_w}$;
        \If {$blk$  $\land$  $(b_q \in x_s)$ $\land$ $(Z_{b_q}=0)$};
        \State $U_w,Z_{b_q},blk, R'_q\gets$ $\Call{FFWavPair}{g,\Lambda_{b_q},c_{gs},l_{q},R_{q},V_{q},U_w}$;
        \ElsIf {$blk$  $\land$ $((b_q \notin x_s)$ $\lor$ $(Z_{b_q}=1))$ $\land$ $(s_u < |S|)$}
        \State $append(x_s, b_q)$;
        \State $s_u \gets s_u + 1$;
        \State $U_w,Z_{b_q},blk, R'_q\gets$ $\Call{FFWavPair}{g,\Lambda_{b_q},c_{gs},l_{q},R_{q},V_{q},U_w}$;
        \If {$blk$};
        \State $x_{s}.pop()$;
        \State $s_u \gets s_u - 1$;
        \EndIf
        \EndIf
        \ElsIf{$(b_q \in x_s)$ $\land$ $(Z_{b_q}=0)$};
        \State $U_w,Z_{b_q},blk, R'_q \gets \Call{FFWavPair}{g,\Lambda_{b_q},c_{gs},l_{q},R_{q},V_{q},U_w}$;
        \ElsIf{$((a_q \notin x_s)$ $\lor$ $(Z_{a_q}=1))$ $\land$ $(s_u < |S|)$};
        \State $append(x_s, a_q)$;
        \State $s_u \gets s_u + 1$;
        \State $U_w,Z_{a_q},blk, R'_q \gets \Call{FFWavPair}{g,\Lambda_{a_q},c_{gs},l_{q},R_{q},V_{q},U_w}$;
        \If {$blk$};
        \State $x_{s}.pop()$;
        \State $s_u \gets s_u - 1$;
        \EndIf
        \Else
        \State $blk \gets true$;
		\EndIf
        \If{$\lnot blk$};
        \State $Q_{acc}\gets Q_{acc}+1$;
        \EndIf
        \EndFor
        \State \textbf{return} $Q_{acc}, s_{u}, x_{s}$;
	\end{algorithmic}
\end{algorithm}


Using OP distribution, the pseudocode to simulate placement of EPPSs, photon routing, and wavelength pair assignment is given in Algorithm 1. For each entanglement demand (step 1), a minimum-loss route is identified (step 2) between the two nodes ($a_q$ and $b_q$) to be entangled. $\textproc{MinLossRoute}()$ calculates all possible routes between $a_q$ and $b_q$ and the losses incurred by them, and returns the route that offers lowest loss between $a_q$ and $b_q$. If one ($a_q$) of the two nodes has an EPPS available, wavelength pair assignment is attempted using first-fit wavelength pair assignment scheme, whose pseudocode is given in Algorithm 2. The first-fit wavelength pair assignment scheme for entangled photon pair distribution is similar to the first-fit wavelength assignment scheme used in classical optical networks \cite{zang2000review} in the sense that all the available wavelengths are indexed sequentially and then during wavelength assignment, the search for available wavelengths is done starting from the first index on all the fiber links along the route (referred to as `wavelength continuity constraint'). However, for distribution of entangled photon pairs that are found at a correlated pair of wavelengths, a wavelength pair is assigned for each entanglement demand instead of single wavelength. Most importantly, in addition to availability of a wavelength pair, whether or not it meets the required ebit rate ($R_q$) and visibility ($V_q$), is also checked (step 7 in Alg. 2). We consider the same (i.e., first-fit wavelength pair assignment) scheme for wavelength pair assignment for all the compared cases in this work to better highlight the impact of EPPS placement and OP/BP distribution. However, more optimal and complex wavelength pair assignment schemes can be designed by interested researchers to further improve spectrum utilization. If a wavelength pair is found, the entangled demand is accepted (step 30, Alg. 1) or else marked as blocked (`$blk$') (returned by $\textproc{FFWavPair}()$). In case it is blocked, another attempt is made by searching for an available EPPS at the second entangling node ($b_q$) (step 5, Alg. 1). If no EPPS is placed at $b_q$ or if the placed EPPSs are fully utilized, another EPPS is placed at node $b_q$ (subject to availability ($s_u < |S|$)) and then another final attempt is made to satisfy the entanglement demand (steps 7-10, Alg. 1). However, only if an entanglement demand is accepted using a new EPPS placement, it remains placed, otherwise revoked (step 12). If no EPPS is found at node $a_q$ (step 3), the same is checked at $b_q$ (step 16) followed by wavelength pair assignment. If none of the two nodes ($a_q$ and $b_q$) have an EPPS placed already (which would be the case for the first few demand arrivals) or if the available EPPSs at $a_q$ are fully utilized (step 18), another EPPS is placed (subject to availability ($s_u < |S|$)) at $a_q$ (step 19) followed by wavelength pair assignment (step 21). In none of the wavelength pairs generated by the available EPPSs could meet the required ebit rate and visibility of entanglement demand or if all the available EPPSs are utilized, the entanglement demand is finally marked as blocked ($`blk'$) (step 27). The number of accepted entanglement demands ($Q_{acc}$), number of EPPSs utilized ($s_u$), and the locations of EPPSs placed ($x_s$) are returned. Though OP distribution scheme simplifies routing and EPPS placement, it requires an EPPS to be placed at at least one of the two entangling nodes. For a network of $N$ number of nodes, to entangle all the node pairs, it requires EPPSs at at least $N-1$ nodes. Thus, we analyze the BP distribution scheme next that relaxes this constraint.

\begin{algorithm}[!t]
	\footnotesize
	\caption*{Algorithm 2: First-Fit Wavelength-Pair Assignment}
	\label{algo:firstfit}
	\begin{algorithmic}[1]
        \Function{FFWavPair}{$g,\Lambda,C,l,R_{th},V_{th},U$}
        \State $blk \gets true$; $count \gets 0$; $z \gets 0$; $R \gets 0$;
		\For{$i\gets$ ($\Lambda-(1+g\bmod2$)) to ($\Lambda-1$)}
        \If {$U_i= 0$};
        \State $R \leftarrow C_{i \bmod g}/l$;
        \State Calculate $V(d)$ using Eq. (5);
        \If{$(R \geq R_{th}) \land (V(d) \geq V_{th})$};
        \State $blk\leftarrow false$;
        \State $U_i=1$; $U_{i+2(\Lambda-i)}=1$;
        \If{$count = (g\bmod2) - 1$};
        \State $z=1$;
        \EndIf
        \State \textbf{break}
        \EndIf
        \Else
        \State $count \gets count + 1$
        \EndIf
        \EndFor
        \State\Return $U,z,blk,R$
        \EndFunction
	\end{algorithmic}
    
\end{algorithm}

The pseudocode to simulate placement of EPPSs, photon pair routing, and wavelength pair assignment using BP distribution scheme is given in Algorithm 3. Here, an EPPS placed at a node other than the two entangling nodes is used to provision each entanglement demand. Thus, for each demand (step 1, Alg. 3), all the nodes except the two nodes to be entangled are searched (steps 3-4) and two minimum-loss routes ($l'_{a_q}$ and $l'_{b_q}$) are calculated from each searched node to the two nodes ($a_q$ and $b_q$) to be entangled. The node that offers minimum total loss ($l$) is selected as the location at which an EPPS is to be placed (if not placed already) to serve this entanglement demand (steps 7-16). Once EPPS location is identified (step 20), wavelength pair assignment is attempted (step 21) for this demand or else blocked otherwise (step 27). If wavelength pair assignment remains unsuccessful (step 21) on placing a new EPPS as well (step 22), this EPPS placement is revoked (step 23).

\begin{algorithm}[!t]
	\footnotesize
	\caption*{Algorithm 3: Both Photons (BP) distribution}
	\label{algo1}
	\textbf{Input:} $N, E, d_e, S, c_{gs}, \Lambda, g,$ $Q\{a_{q},b_{q},V_{q},R_{q}\}$, $s_{u}\gets 0$, $U_w$, $Z_s$, $Q_{acc}\gets 0$,  $l_{q}\gets \Theta$,  $y\gets 0$,  $x_s\gets 0\text{~} \forall s$\\
	\textbf{Output:} $Q_{acc}, s_{u}, x_{s}$
	\begin{algorithmic}[1]
		\For{$q\gets$ $1$ to $|Q|$}
        \State $place \gets false$;
        \For{$t\gets 1$ to $|N|$}
        \If{$(a_q \neq N_t)$ $\land$ $(b_q \neq N_t)$};
        \State $l'_{a_q} \gets \Call{MinLossRoute}{a_q,N_t,N,E,d_e}$;
        \State $l'_{b_q} \gets \Call{MinLossRoute}{b_q,N_t,N,E,d_e}$;
        \State $l \gets l'_{a_q} + l'_{b_q}$
        \If{$(l<l_q)$ $\land$ $(x_t \neq 0)$ $\land$ $(Z_t = 0)$};
        \State $l_q\gets l$;
        \State $y\gets t$;
        \ElsIf{$(l<l_q)$ $\land$ $((x_t=0) \lor (Z_t = 1))$ $\land$ $(s_u<|S|)$};
        \State $append(x_s, N_t)$;
        \State $s_u\gets s_u + 1$;
        \State $l_q\gets l$;
        \State $y\gets t$;
        \State $place\gets true$;
        \EndIf
        \EndIf
        \EndFor
        \If {$y\neq0$};
        \State $U_w,Z_y,blk, R'_q \gets \Call{FFWavPair}{g,\Lambda_{y},c_{vy},l_{q},R_{q},V_{q},U_w}$;
        \If {$blk$ $\land$ $place$};
        \State $x_{s}.pop()$;
        \State $s_u \gets s_u - 1$;
        \EndIf
        \Else
        \State $blk \gets true$;
		\EndIf
        \If{$\lnot blk$};
        \State $Q_{acc}\gets Q_{acc}+1$;
        \EndIf
        \EndFor
        \State \textbf{return} $Q_{acc}, s_{u}, x_{s}$
	\end{algorithmic}
\end{algorithm}

BP distribution scheme can be employed even with a single EPPS in a network if the EPPS generates sufficient rates and wavelength pairs. However, transmitting both the photons of each pair to farther node pairs from the EPPS location leads to higher losses. Thus, both OP and BP distribution schemes have pros and cons. Hence, we also study a hybrid distribution scheme referred to as `any photon' (AP) in this work, where entanglement demand provisioning is first attempted using BP (Alg. 3), however, if the entanglement demand is blocked, another attempt is made to provision it using OP (Alg. 1). AP provides flexibility in EPPS placement by relaxing the constraints `$N-1$ nodes must have EPPS/s in a network of $N$ nodes' and `an EPPS at nodes other than the two entangling nodes must be used' imposed in case of OP and BP, respectively. We evaluate the performance of OP, BP, and AP distribution schemes in the next section and observe the EPPS placement done once all the entanglement demands are served (accepted and blocked). Based on the observations drawn, a further analysis of entanglement distribution is performed with pre-placement of all the available EPPSs simultaneously at different node/s in the network, as explained in detail in the next Section.

\section{Performance Evaluation and Discussion}

\begin{figure*}[t]
\centering
\vspace{-0.2em}
\begin{subfigure}[b]{0.16\textwidth}
  \centering
  \includegraphics[width=\textwidth]{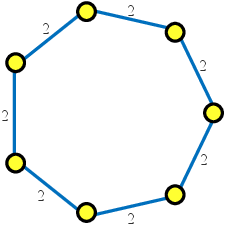}
  \vspace{-1em}
  \caption{7-RING}
\end{subfigure}\hfill
\begin{subfigure}[b]{0.16\textwidth}
  \centering
  \includegraphics[width=\textwidth]{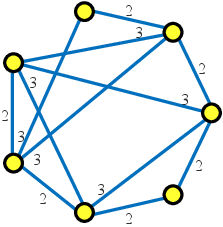}
  \vspace{-1em}
  \caption{7-PMESH}
\end{subfigure}\hfill
\begin{subfigure}[b]{0.16\textwidth}
  \centering
  \includegraphics[width=\textwidth]{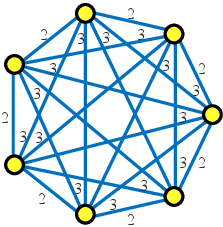}
  \vspace{-1em}
  \caption{7-FMESH}
\end{subfigure}\hfill
\begin{subfigure}[b]{0.16\textwidth}
  \centering
  \includegraphics[width=\textwidth]{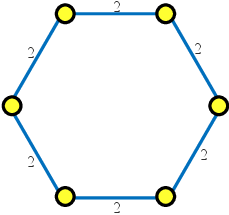}
  \vspace{-1em}
  \caption{6-RING}
\end{subfigure}\hfill
\begin{subfigure}[b]{0.16\textwidth}
  \centering
  \includegraphics[width=\textwidth]{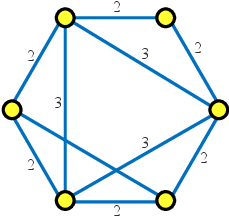}
  \vspace{-1em}
  \caption{6-PMESH}
\end{subfigure}\hfill
\begin{subfigure}[b]{0.16\textwidth}
  \centering
  \includegraphics[width=\textwidth]{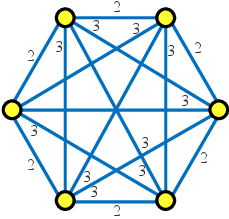}
  \vspace{-1em}
  \caption{6-FMESH}
\end{subfigure}
\vspace{-0.3em}
\caption{7-node and 6-node ring, full mesh (FMESH), and partial mesh (PMESH) random (generated using Erdős--Rényi graph model \cite{erdos1959random}) network topologies. The values shown at edges are link distances.}
\label{fig:wattstopo}
\vspace{-1.5em}
\end{figure*}

To evaluate the performance of WDM-based distributed entanglement distribution in quantum optical networks with multiple EPPSs using the OP, BP, and AP distribution schemes described in the previous section, we consider various network topologies shown in Fig. \ref{fig:wattstopo}. Here, the network topology size is considered based on the number of wavelength pairs generated by multiple EPPSs. With $|S|$ EPPSs generating $p$ ($= (g-1)/2$) wavelength pairs, to entangle all the node pairs in a network of $|N|$ nodes, $|S|\cdot p =$ $|N|\choose2$. The total number of wavelength pairs obtained in C-band from the wavelength-multiplexed spectra of six 3WPSs (see Fig. \ref{fig:multisource_spectra}) and four 5WPSs is 18 and 20, respectively. Thus, we consider 6-node and 7-node network topologies (shown in Fig. \ref{fig:wattstopo}) for performance analyses, where $6\choose2$$=15$ and $7\choose2$$=21$ unique node pairs (that are to be entangled) are slightly lower and higher, respectively, than the total available wavelength pairs (18 and 20). However, denser networks (i.e., higher number of nodes) can be considered for this study when a higher number of wavelength pairs are available (such as using finer wavelength granularity). To generate random partial mesh network topologies shown in Fig. \ref{fig:wattstopo}, we use the Erdős--Rényi \cite{erdos1959random} graph model. The link-lengths are considered based on the available EPPS rates and repeaterless entanglement distribution studied in this work. However, longer links can be simulated when higher EPPS rates and/or quantum repeaters are considered. We use \emph{networkx} (a python library) to generate and simulate all the graph-related tasks in this work. Analyses for different network topologies is done in this work to identify if the comparative performance of OP, BP, and AP distribution schemes vary with network topologies. Thus, ring, partial-mesh, and full-mesh network topologies are considered. By varying the edge probability $0<\mathbb{P}_{e}<1$ in Erdős--Rényi graph $G_{e}(|N|,\mathbb{P}_{e})$, we analyzed WDM-based entanglement distribution for several random partial mesh network topologies and we present simulation results for two instances of partial mesh network topologies of 7- and 6- nodes, shown in Fig. \ref{fig:wattstopo}(b) and Fig. \ref{fig:wattstopo}(e), respectively. The simulation results presented in this section have been averaged for $10^6$ simulation runs for each variation. The simulation parameters are given in Table \ref{tab3}.

\begin{table}[!t]
\scriptsize
\centering
\caption{Simulation parameters}
\vspace{-0.7em}
\label{tab3}
\begin{tabular}{p{4.23cm}p{3.8cm}}
\hline
\hline
\multicolumn{1}{l}{Parameter} & \multicolumn{1}{l}{Value} \\ \hline
Set of demanded ebit rates & $\{100, 500, 1000, 2000\} (ebps)$ \\
\\[-1em]
Set of demanded entanglement visibility & $\{71, 75, 80, 85, 90\}$ (\%)\\
\\[-1em]
Fiber attenuation ($\alpha$) & 0.2 (dB/km) \\
\\[-1em]
WSS loss ($L_{wss}$) & $3.5$ (dB) \\
\\[-1em]
OC loss ($L_{oc}$) & $0.5$ (dB) \\
\\[-1em]
$N_{acc}$ & $20$ (cps) \\
\\[-1em]
Tuple of ebit-rate generated using 3WPS & $\{21.43, 29.65, 33.83\}\times10^3$ (ebps)\\
\\[-1em]
Tuple of visibility for 3WPS & $\{85, 87, 93\}$ (\%)\\
\\[-1em]
Tuple of ebit-rate generated using 5WPS & $\{42.62, 73.34, 34.53, 76.11, 42.79\}$ $\times10^3$ (ebps)\\
\\[-1em]
Visibility for all wavelength pairs of 5WPS & $92$ (\%)\\
\hline
\hline
\end{tabular}
\end{table}

\begin{figure}[!t]
\centering
\begin{subfigure}[b]{0.445\textwidth}
  \centering
  \includegraphics[width=\textwidth]{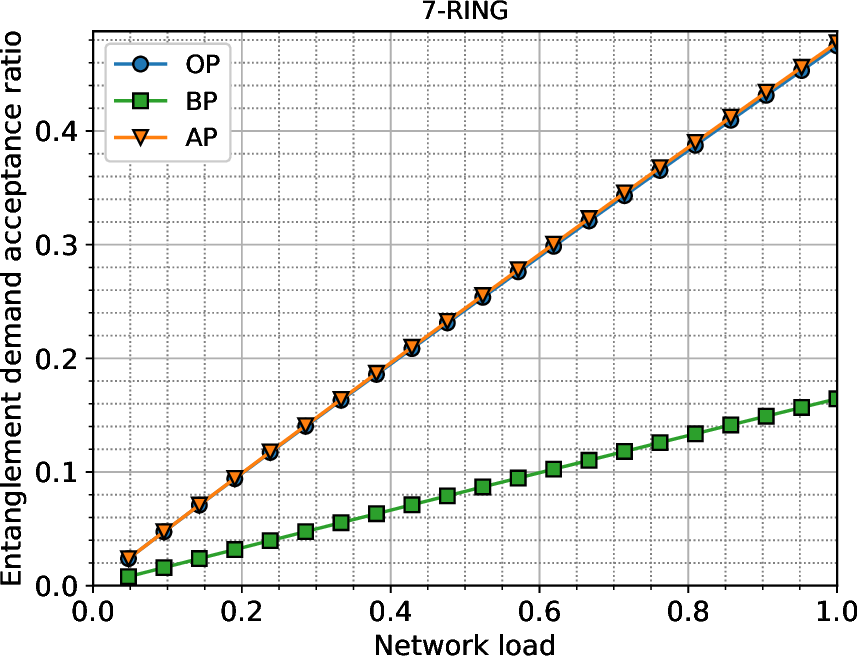}
  \vspace{-1.5em}
  \caption{}
\end{subfigure}\hfill
\begin{subfigure}[b]{0.445\textwidth}
  \centering
  \vspace{1em}
  \includegraphics[width=\textwidth]{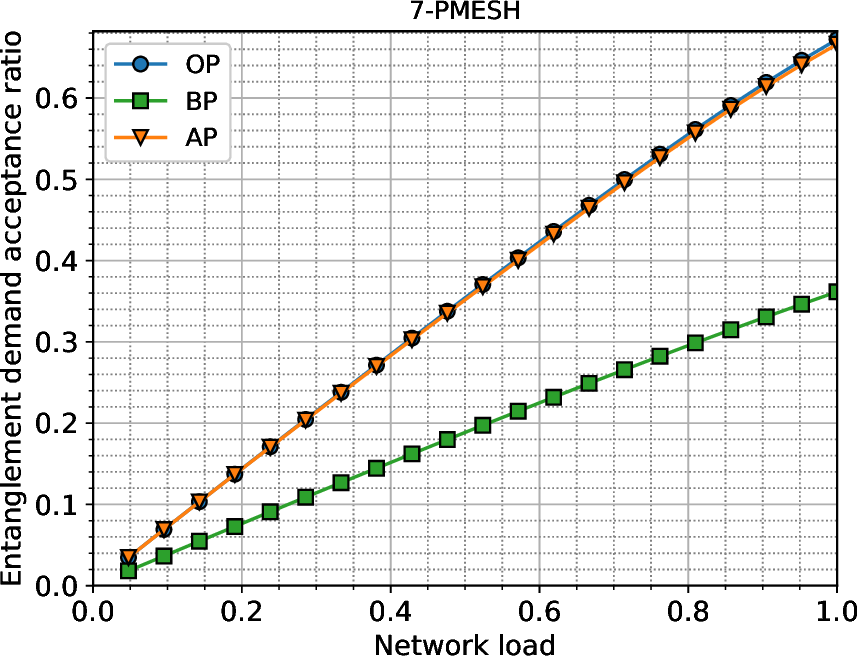}
  \vspace{-1.5em}
  \caption{}
\end{subfigure}\hfill
\begin{subfigure}[b]{0.445\textwidth}
  \centering
  \vspace{1em}
  \includegraphics[width=\textwidth]{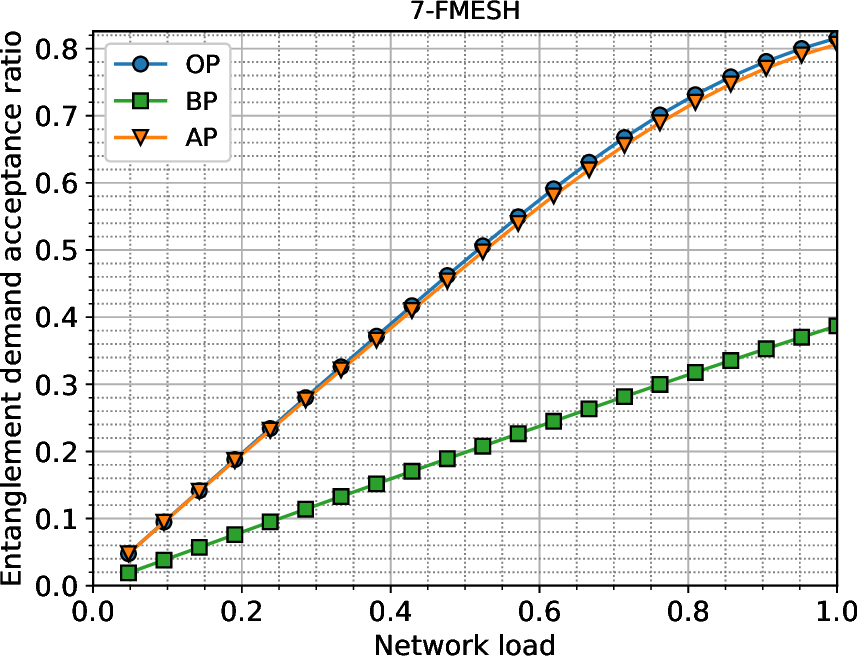}
  \vspace{-1.5em}
  \caption{}
\end{subfigure}
\caption{Entanglement demand acceptance ratio (EDAR) versus network load (normalized) obtained using OP, BP, and AP distribution schemes with six 3WPSs for (a) 7-RING network (see Fig. \ref{fig:wattstopo}(a)), (b) 7-PMESH network (see Fig. \ref{fig:wattstopo}(b)), and (c) 7-FMESH network (see Fig. \ref{fig:wattstopo}(c)).}
\label{fig:EDAR_7NODE_3WPS}
\end{figure}

\begin{figure}[!t]
\centering
\begin{subfigure}[b]{0.45\textwidth}
  \centering
  \includegraphics[width=\textwidth]{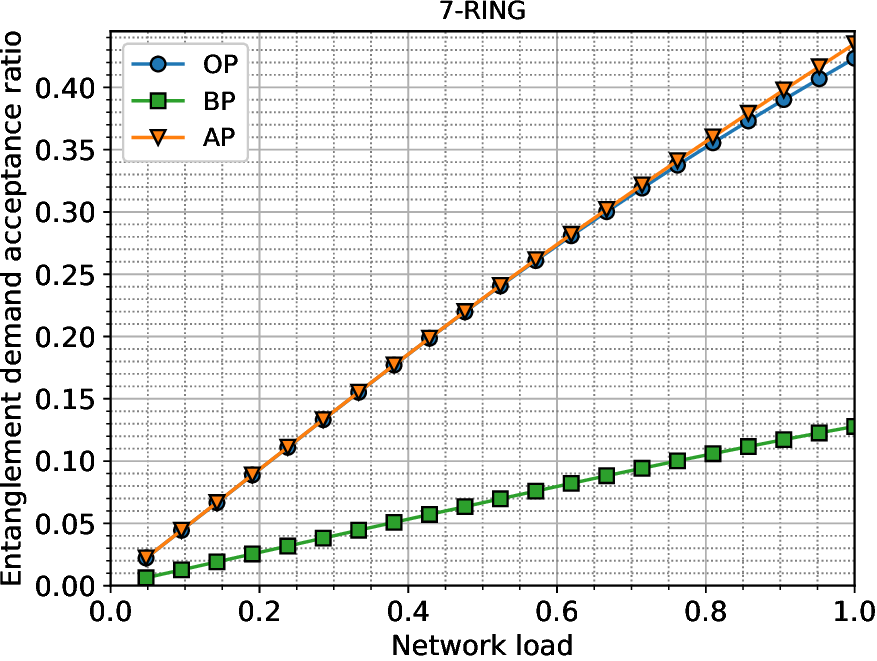}
  \vspace{-1.5em}
  \caption{}
\end{subfigure}\hfill
\begin{subfigure}[b]{0.45\textwidth}
  \centering
  \vspace{1em}
  \includegraphics[width=\textwidth]{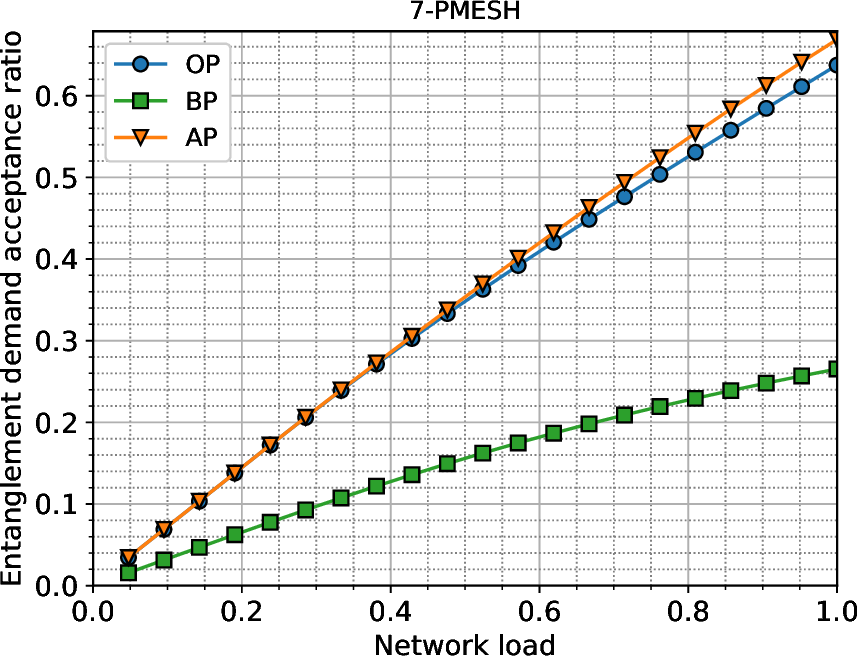}
  \vspace{-1.5em}
  \caption{}
\end{subfigure}\hfill
\begin{subfigure}[b]{0.45\textwidth}
  \centering
  \vspace{1em}
  \includegraphics[width=\textwidth]{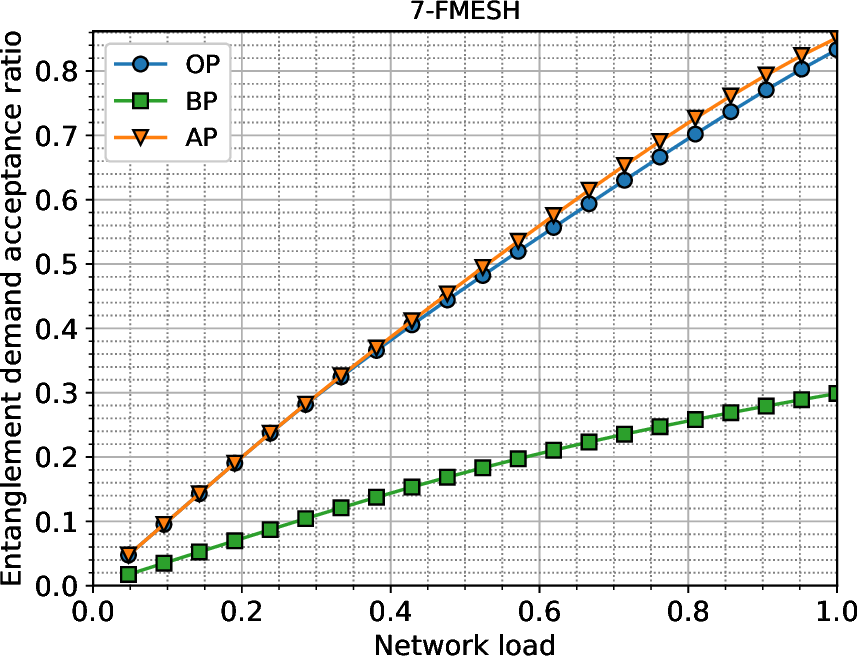}
  \vspace{-1.5em}
  \caption{}
\end{subfigure}
\caption{Entanglement demand acceptance ratio (EDAR) versus network load (normalized) obtained using OP, BP, and AP distribution schemes with four 5WPSs for (a) 7-RING network, (b) 7-PMESH network, and (c) 7-FMESH network.}
\label{fig:EDAR_7NODE_5WPS}
\end{figure}

As the entanglement demands arrive incrementally in the network, EPPS placement, photon routing, wavelength pair assignment is performed for the six network topologies shown in Fig. \ref{fig:wattstopo}. The effect of photon pair loss and visibility degradation (using Eq. (5)) is estimated to ensure that the minimum required ebit rate and visibility by each demand are met (step 7, Alg. 2). We assume that polarization auto-compensators at the entanglement receivers can recover any depolarization that the photon pairs undergo en-route the network. We calculate entanglement demand acceptance ratio (EDAR) as the ratio of the number of entanglement demands accepted to the number of entanglement demands arrived in the network. EDAR indicates the impact on network performance with combined variation of EPPS placement and OP/BP/AP distribution schemes using a given (fixed) number of resources (EPPSs, wavelength pairs generated, photon pairs count at each wavelength pair, nodes and links in network topology). A higher value of EDAR signifies better EPPS placement and photon distribution scheme (OP/BP/AP). We calculate EDAR as the network load increases (i.e., as the number of entanglement demands arrive in the network). The network load represented in Figs. \ref{fig:EDAR_7NODE_3WPS}-\ref{fig:EDAR_7NODE_5WPS} is normalized by the maximum number of unique node pairs in the simulated network topology. In Fig. \ref{fig:EDAR_7NODE_3WPS}, EDAR obtained using OP, BP, and AP distribution schemes with six 3WPSs is shown for 7-node ring (7-RING), 7-node partial mesh (7-PMESH), and 7-node full mesh (7-FMESH) topologies (shown in Figs. \ref{fig:wattstopo}(a-c)). It is observed that as the network connectivity increases (ring $\rightarrow$ partial-mesh $\rightarrow$ full-mesh), EDAR increases using any of the OP, BP, and AP distribution schemes. This is because with increasing mesh connectivity, shorter routes (i.e., with fewer hops and lesser link-distances) with lower losses get selected, enabling entanglement demands with higher ebit rate and/or visibility getting accepted. However, the comparative performance of OP, BP, and AP remains similar, where OP and AP achieve almost same EDAR as the network load varies and BP achieves significantly lower EDAR as compared to OP and AP for all the three network topologies.

\begin{figure}[!t]
\centering
\begin{subfigure}[b]{0.45\textwidth}
  \centering
  \includegraphics[width=\textwidth]{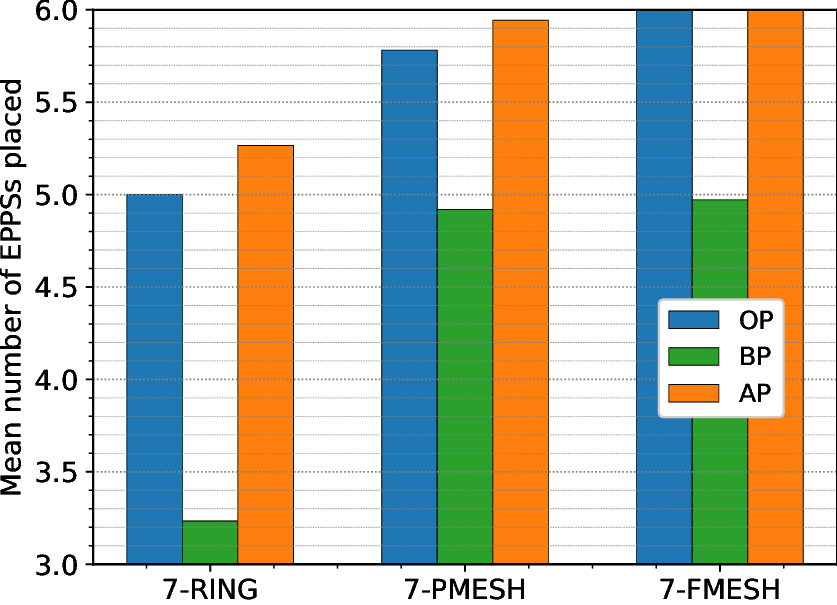}
  \vspace{-1.5em}
  \caption{}
\end{subfigure}\hfill
\begin{subfigure}[b]{0.45\textwidth}
  \centering
  \includegraphics[width=\textwidth]{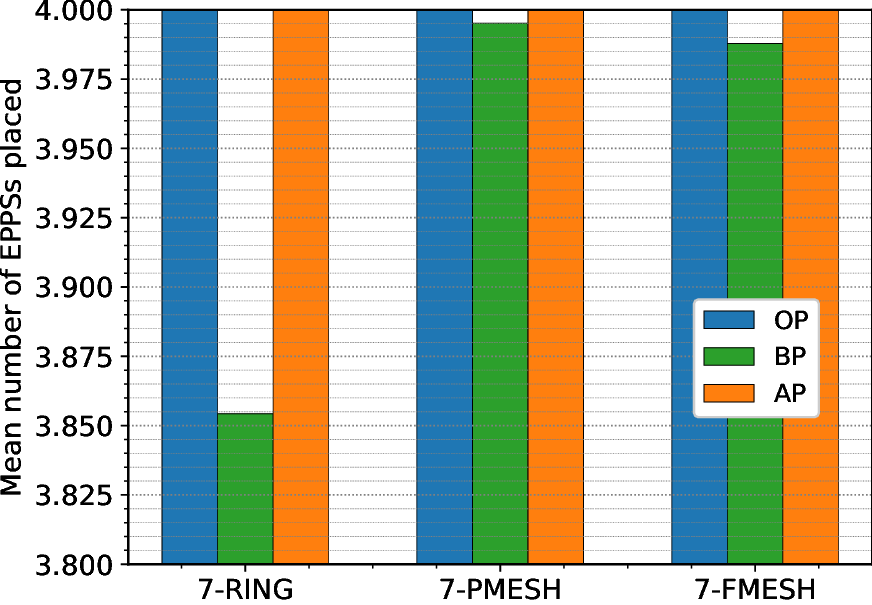}
  \vspace{-1.5em}
  \caption{}
\end{subfigure}
\caption{Mean number of (a) 3WPSs and (b) 5WPSs placed using OP, BP, and AP schemes in the 7-RING, 7-PMESH, and 7-FMESH networks.}
\label{fig:EPPP_7NODE}
\end{figure}

When using 5WPSs, AP performs better than OP at higher network loads, as shown in Fig. \ref{fig:EDAR_7NODE_5WPS}, since with four 5WPSs in a 7-node network, OP fails to entangle some node pairs as it requires an EPPS at at least 6 (i.e., $|N|-1$) nodes. Thus, AP is advantageous when using broader-band EPPSs such as 5WPS as compared to 3WPS in this case. BP still performs the worst for all the three network topologies, as shown in Fig. \ref{fig:EDAR_7NODE_5WPS}. We analyzed and observed similar comparative performance (not shown here for brevity) of OP, BP, and AP distribution schemes for all the three 6-node network topologies (Figs. \ref{fig:wattstopo}(d-f)). To observe EPPS placement and to determine the reasons behind the obtained comparative performance of OP, BP, and AP schemes, we plot the mean number of EPPSs placed in each considered case in Fig. \ref{fig:EPPP_7NODE}. Since an EPPS is placed only when an entanglement demand cannot be provisioned using existing EPPSs (if any) in the network and if a new EPPS placement (subject to availability) can provision it (refer Algs. 1 and 3), we observe from Fig. \ref{fig:EPPP_7NODE} that using BP distribution scheme, 100\% EPPS placement could not be achieved in all of the considered cases. This is because BP always selects a node other than the two entangling nodes for EPPS placement in which case both the photons are routed out of the source node and undergo higher loss (due to an additional WSS (see Fig. \ref{fig:nodearch})) than OP (where one photon is sent to the entanglement receiver within the source node (see Fig. \ref{fig:nodearch})). This indicates that the ebit rate and/or visibility required by certain entanglement demands could not be met by placing EPPSs (3WPS/5WPS) at any of the nodes other than the entangling node pair, but could be met when EPPSs were placed at one of the two entangling nodes (i.e., using OP). This is evident from Fig. \ref{fig:EPPP_7NODE}(b) where, using OP, all the four 5WPSs were placed for all the three network topologies. From Fig. \ref{fig:EPPP_7NODE}(a), it is observed that EPPS placement, when using OP, increases with increase in network connectivity (ring $\rightarrow$ partial-mesh $\rightarrow$ full-mesh) indicating that for ring and partial mesh topologies, ebit rate and/or visibility for certain entanglement demands could not be met due to longer routes among node pairs in case of ring and partial-mesh network topologies. However, in case of full-mesh, all the six 3WPSs were placed when using OP since full-mesh provides direct connectivity among all the node pairs. Since AP attempts entanglement distribution using first BP and then OP, it enables flexibility to place different EPPSs (six 3WPSs and four 5WPSs) at any node/s in the network while achieving similar EDAR as that of OP when six using 3WPSs, and better EDAR than that of OP when using four 5WPSs.

Based on the incremental EPPS placements done as the entanglement demands arrive in the network, we observed that the final EPPS placements varied both using 3WPSs and 5WPSs for all the considered network topologies. It is also observed that when EPPSs were placed at nodes of higher degree (i.e., nodes with higher number of links connected to it), better EDAR was achieved, however, it was not the case when multiple EPPSs got placed at those nodes. Since EPPS placement order was random based on the random arrival order of entanglement demands in this case, a few EPPSs placed at non-optimal node locations could not be fully utilized. Thus, to identify the effect of EPPS placement on entanglement distribution, we now consider pre-placement of given EPPSs (i.e., six 3WPSs and four 5WPSs) and analyze different ways of placement. Since AP distribution scheme enables flexibility in EPPS placement while achieving similar or better performance than OP, we consider AP distribution for further analysis on EPPS pre-placement variation. Since the comparative performance of OP, BP, and AP does not vary with network connectivity in both the cases of 3WPSs and 5WPSs, we now consider partial mesh (which represent more realistic optical networks) 7-node and 6-node network topologies shown in Figs. \ref{fig:wattstopo}(b) and \ref{fig:wattstopo}(e) for further analysis.

\begin{figure*}[!t]
\centering
\vspace{-0.2em}
\begin{subfigure}[b]{0.19\textwidth}
  \centering
  \includegraphics[width=\textwidth]{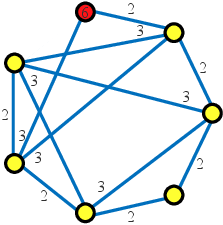}
  \vspace{-1em}
  \caption{7N-3WPS-P1}
\end{subfigure}\hfill
\begin{subfigure}[b]{0.19\textwidth}
  \centering
  \includegraphics[width=\textwidth]{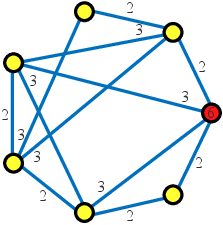}
  \vspace{-1em}
  \caption{7N-3WPS-P2}
\end{subfigure}\hfill
\begin{subfigure}[b]{0.19\textwidth}
  \centering
  \includegraphics[width=\textwidth]{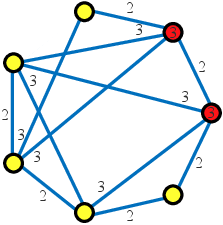}
  \vspace{-1em}
  \caption{7N-3WPS-P3}
\end{subfigure}\hfill
\begin{subfigure}[b]{0.19\textwidth}
  \centering
  \includegraphics[width=\textwidth]{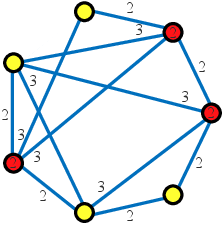}
  \vspace{-1em}
  \caption{7N-3WPS-P4}
\end{subfigure}\hfill
\begin{subfigure}[b]{0.19\textwidth}
  \centering
  \includegraphics[width=\textwidth]{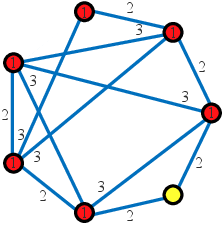}
  \vspace{-1em}
  \caption{7N-3WPS-P5}
\end{subfigure}
\vspace{-0.3em}
\caption{Different placements (P1, P2, P3, P4, P5) of 3WPSs in 7-PMESH network based on betweenness centrality ($C_{bw}$) and dominant sets (of size $m$): (a) $C_{bw}^{min}$, (b) $C_{bw}^{max}$, (c) $m=2$, (d) $m=3$, (e) $m=6$. The value/s specified inside node/s (circle/s) is the number of 3WPSs placed at that node.}
\label{fig:place3WPS_7}
\vspace{-0.9em}
\end{figure*}

\begin{figure*}[!t]
\centering
\vspace{-0.2em}
\begin{subfigure}[b]{0.2\textwidth}
  \centering
  \includegraphics[width=\textwidth]{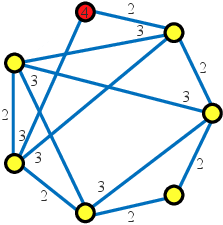}
  \vspace{-1em}
  \caption{7N-5WPS-P1}
\end{subfigure}\hfill
\begin{subfigure}[b]{0.2\textwidth}
  \centering
  \includegraphics[width=\textwidth]{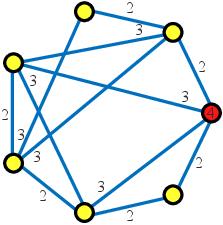}
  \vspace{-1em}
  \caption{7N-5WPS-P2}
\end{subfigure}\hfill
\begin{subfigure}[b]{0.2\textwidth}
  \centering
  \includegraphics[width=\textwidth]{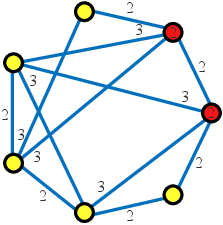}
  \vspace{-1em}
  \caption{7N-5WPS-P3}
\end{subfigure}\hfill
\begin{subfigure}[b]{0.2\textwidth}
  \centering
  \includegraphics[width=\textwidth]{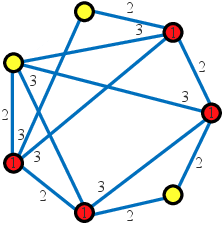}
  \vspace{-1em}
  \caption{7N-5WPS-P4}
\end{subfigure}
\vspace{-0.3em}
\caption{Different placements (P1, P2, P3, P4) of 5WPSs in 7-PMESH network based on betweenness centrality ($C_{bw}$) and dominant sets (of size $m$): (a) $C_{bw}^{min}$, (b) $C_{bw}^{max}$, (c) $m=2$, (d) $m=4$. The value/s specified inside node/s (circle/s) is the number of 5WPSs placed at that node.}
\label{fig:place5WPS_7}
\vspace{-0.9em}
\end{figure*}

\begin{figure*}[!t]
\centering
\vspace{-0.2em}
\begin{subfigure}[b]{0.19\textwidth}
  \centering
  \includegraphics[width=\textwidth]{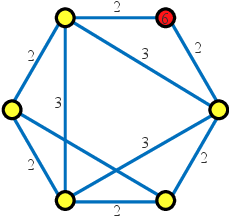}
  \vspace{-1em}
  \caption{6N-3WPS-P1}
\end{subfigure}\hfill
\begin{subfigure}[b]{0.19\textwidth}
  \centering
  \includegraphics[width=\textwidth]{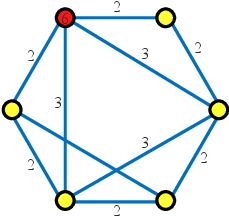}
  \vspace{-1em}
  \caption{6N-3WPS-P2}
\end{subfigure}\hfill
\begin{subfigure}[b]{0.19\textwidth}
  \centering
  \includegraphics[width=\textwidth]{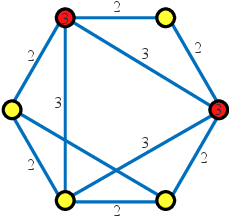}
  \vspace{-1em}
  \caption{6N-3WPS-P3}
\end{subfigure}\hfill
\begin{subfigure}[b]{0.19\textwidth}
  \centering
  \includegraphics[width=\textwidth]{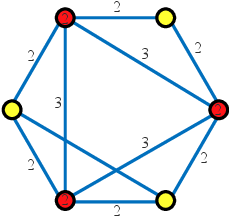}
  \vspace{-1em}
  \caption{6N-3WPS-P4}
\end{subfigure}\hfill
\begin{subfigure}[b]{0.19\textwidth}
  \centering
  \includegraphics[width=\textwidth]{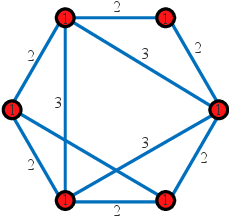}
  \vspace{-1em}
  \caption{6N-3WPS-P5}
\end{subfigure}
\vspace{-0.3em}
\caption{Different placements (P1, P2, P3, P4, P5) of 3WPSs in 6-PMESH network based on betweenness centrality ($C_{bw}$) and dominant sets (of size $m$): (a) $C_{bw}^{min}$, (b) $C_{bw}^{max}$, (c) $m=2$, (d) $m=3$, (e) one 3WPS at each node. The value/s specified inside node/s (circle/s) is the number of 3WPSs placed at that node.}
\label{fig:place3WPS_6}
\vspace{-0.9em}
\end{figure*}

\begin{figure*}[!t]
\centering
\vspace{-0.2em}
\begin{subfigure}[b]{0.2\textwidth}
  \centering
  \includegraphics[width=\textwidth]{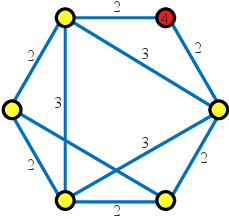}
  \vspace{-1em}
  \caption{6N-5WPS-P1}
\end{subfigure}\hfill
\begin{subfigure}[b]{0.2\textwidth}
  \centering
  \includegraphics[width=\textwidth]{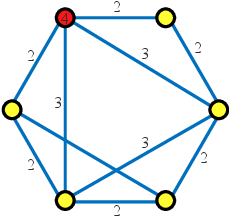}
  \vspace{-1em}
  \caption{6N-5WPS-P2}
\end{subfigure}\hfill
\begin{subfigure}[b]{0.2\textwidth}
  \centering
  \includegraphics[width=\textwidth]{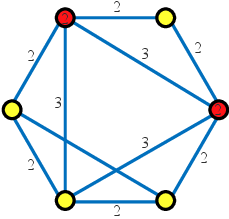}
  \vspace{-1em}
  \caption{6N-5WPS-P3}
\end{subfigure}\hfill
\begin{subfigure}[b]{0.2\textwidth}
  \centering
  \includegraphics[width=\textwidth]{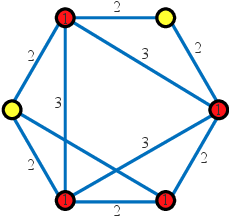}
  \vspace{-1em}
  \caption{6N-5WPS-P4}
\end{subfigure}
\vspace{-0.3em}
\caption{Different placements (P1, P2, P3, P4) of 5WPSs in 6-PMESH network based on betweenness centrality ($C_{bw}$) and dominant sets (of size $m$): (a) $C_{bw}^{min}$, (b) $C_{bw}^{max}$, (c) $m=2$, (d) $m=4$. The value/s specified inside node/s (circle/s) is the number of 5WPSs placed at that node.}
\label{fig:place5WPS_6}
\vspace{-1em}
\end{figure*}

We leverage the classical graph theory concepts of `betweenness centrality' and `dominant sets' to analyze different EPPS placement patterns shown in Figs. \ref{fig:place3WPS_7}-\ref{fig:place5WPS_6}. Betweenness centrality of a node in a network is a measure of how many shortest paths between all the node pairs in the network traverse that node. Since each considered EPPS in this work generate multiple wavelength pairs that are to be routed to multiple node pairs in the network, an EPPS placed at a node with maximum betweenness centrality would be able to transmit entangled photon pairs with lowest losses due to shorter distances between that node and the other nodes in the network. Thus, we leverage the concept of `betweenness centrality' for EPPS placement in this work. Mathematically, betweenness centrality ($C_{bw}$) of a node $v$ in a network of $|N|$ nodes is defined as
\[
C_{bw}(v) = \sum_{\substack{s,t \in N \\ s \neq t, s \neq v, t \neq v}}
\frac{\sigma_{st}(v)}{\sigma_{st}},
\]
where, $\sigma_{st}$ is the number of shortest paths for node pair $s-t$ and $\sigma_{st}(v)$ is the number of those shortest paths that pass through node $v$ (with $v \neq s,t$).

Let $\operatorname{dist}(u,v)$ be the length of a shortest path between nodes $ u, v \in N $ and let $\beta$ ($\ge 1$) be an integer. A subset $ D \subseteq N $ of nodes in a network is called a `distance-$\beta$ dominating set' if $\forall u \in N,\ \exists v \in D \text{ such that } \operatorname{dist}(u,v) \le \beta$. For instance, a distance-$1$ dominating set in an unweighted graph (i.e., $d_e =1, $~$ \forall e\in E$) identifies a set $D (\subseteq N)$ of minimum number of nodes from which all the other nodes in the graph are directly connected by an edge (i.e., one hop away since $\beta=1$). Thus, dominating set is a subset of nodes in a network that are closest with the remaining nodes in that network. Since we aim to analyze the impact of different EPPS placements, we leverage the concepts of `betweenness centrality' and `dominant set' to identify such node/s for EPPS placement that are most-connected and closest to all the other nodes in the network to which entangled photon pairs are to be distributed. When the number of nodes ($=m$) in the subset $|D|$ is specified, it is referred to as a `dominating set of size $m$', where $m$ is an integer such that $ 1 \le m \le |N| $.

We obtain different placements of 3WPSs and 5WPSs in 7-PMESH and 6-PMESH network topologies based on $C_{bw}$ and dominant set of size $m$, as shown in Figs. \ref{fig:place3WPS_7}-\ref{fig:place5WPS_6}. In Fig. \ref{fig:place3WPS_7}(a), all the six 3WPSs are placed at a node with minimum $C_{bw}$ ($C_{bw}^{min}$). In Fig. \ref{fig:place3WPS_7}(b), all the six 3WPSs are placed at a node with maximum $C_{bw}$ ($C_{bw}^{max}$). When multiple nodes have the same $C_{bw}^{min}$ or $C_{bw}^{max}$, one of them has been chosen randomly. Placing all the 3WPSs at a single node in the network would enforce BP distribution (which resulted in lowest EDAR (see Figs. \ref{fig:EDAR_7NODE_3WPS}-\ref{fig:EDAR_7NODE_5WPS})) for most of the node pairs since OP requires an EPPS at one of the two entangling nodes. Thus, we perform distributed placement of six 3WPSs in Figs. \ref{fig:place3WPS_7}(c-e) based on dominant set. In Fig. \ref{fig:place3WPS_7}(c), a dominant set of size $m=2$ (which is also the minimum dominant set for 7-PMESH network) is obtained and three 3WPSs are placed at each of the two (since $m=2$) nodes in the obtained dominant set. Similarly, we obtain more distributed placements by increasing the values of $m$ in Fig. \ref{fig:place3WPS_7}(d) ($m=3$) and Fig. \ref{fig:place3WPS_7}(e) ($m=6$) with two 3WPSs and one 3WPS, respectively, placed at each node in the identified dominant sets. In Fig. \ref{fig:place5WPS_7}, different placements of four 5WPSs in 7-PMESH network topology are obtained similarly. In Fig. \ref{fig:place3WPS_6}, placement of 3WPSs in the 6-PMESH network (shown in Fig. \ref{fig:wattstopo}(e)) is obtained similarly based on $C_{bw}$ and dominant sets of different $m$. A dominant set $D$ of size $m=|N|$ is the same as the original set $N$ of nodes. Thus, with $m=6$, one 3WPS is placed at all the six nodes, as shown in Fig. \ref{fig:place3WPS_6}(e). Different placements of given four 5WPSs in the 6-PMESH network topology are obtained similarly, as shown in Fig. \ref{fig:place5WPS_6}. Since dominant sets are non-unique (i.e., there can be multiple dominant sets of size $m$ for a given graph), the dominant set with maximum sum of $C_{bw}$ of the nodes in the dominant set are selected for EPPS placements in Figs. \ref{fig:place3WPS_7}-\ref{fig:place5WPS_6}.

\begin{figure}[t]
	\centering
	\includegraphics[width=0.9\linewidth]{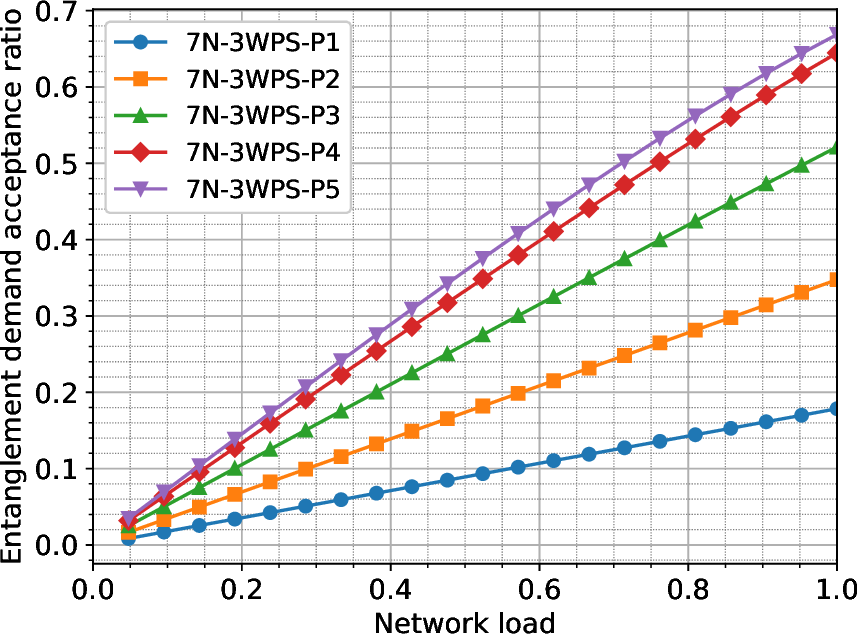}
	\caption{EDAR versus network load for different 3WPS placements represented in Fig. 10 for 7-PMESH network.}
	\label{fig:EDAR_place_3WPS_7PMESH}
\end{figure}

\begin{figure}[t]
	\centering
	\includegraphics[width=0.9\linewidth]{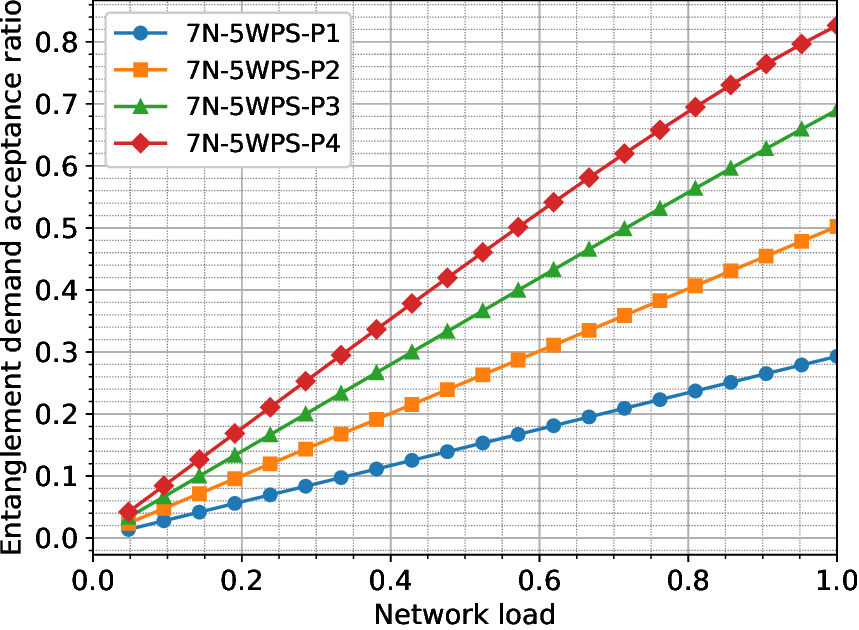}
	\caption{EDAR versus network load for different 5WPS placements represented in Fig. 11 for 7-PMESH network.}
	\label{fig:EDAR_place_5WPS_7PMESH}
\end{figure}

We now analyze the impact of different EPPS placements obtained in Figs. \ref{fig:place3WPS_7}-\ref{fig:place5WPS_6} using AP distribution scheme in terms of EDAR. In Fig. \ref{fig:EDAR_place_3WPS_7PMESH}, EDAR obtained for different 3WPS placements in 7-PMESH network (shown in Fig. \ref{fig:place3WPS_7}) is plotted. It is observed that when all the six 3WPSs are placed at a single node, placing them at a node with $C_{bw}^{max}$ (i.e., placement `7N-3WPS-P2' shown in Fig. \ref{fig:place3WPS_7}(b)) results in higher EDAR as compared to placing them at a node with $C_{bw}^{min}$. This indicates that betweenness centrality is a justified measure to identify suitable node location/s for EPPS placement in quantum networks. Thus, we selected the dominant sets with $C_{bw}^{max}$ among other possible dominant sets of same $m$ in Figs. \ref{fig:place3WPS_7}-\ref{fig:place5WPS_6}. Though $C_{bw}$ is a good measure to identify a suitable single node for EPPS placement, it is observed from Fig. \ref{fig:EDAR_place_3WPS_7PMESH} that with increasingly distributed placement (i.e., with higher $m$) of the given six 3WPSs, EDAR improves. For the most distributed ($m=6$) placement `7N-3WPS-P5' shown in Fig. \ref{fig:place3WPS_7}(e),  $3.75\%$, $28.26\%$, $92.44\%$, and $274.71\%$ higher EDAR has been achieved as compared to `7N-3WPS-P4', `7N-3WPS-P3', `7N-3WPS-P2', and `7N-3WPS-P1' placements, respectively. Such extent of distributability is, however, dependent on how broad-/narrow-band are the EPPSs to be used, how many such EPPSs can be multiplexed within the available optical band (C-band in this work), wavelength granularity, and the network size. For instance, with broader-band EPPS such as the 5WPS, four of which could be wavelength-multiplexed within C-band (as explained in Section IV), $m=4$ is the maximum, as shown in Fig. \ref{fig:place5WPS_7}(d). Nevertheless, the improvement in EDAR with increase in $C_{bw}$ (when all 5WPSs are placed at a single node) and $m$ (when 5WPSs are placed at multiple nodes distributively) persists, as shown in Fig. \ref{fig:EDAR_place_5WPS_7PMESH}. In this case, $19.72\%$, $64.51\%$, and $182.08\%$ higher EDAR has been achieved for `7N-5WPS-P4' placement as compared to `7N-5WPS-P3', `7N-5WPS-P2', and `7N-5WPS-P1' placements, respectively. Thus, depending on the bandwidth and wavelength-granularity of EPPSs, maximum possible distributive placement of EPPSs is advantageous. For instance, when a single EPPS that generates $|N|\choose2$ wavelength pairs is used to entangle all the $|N|\choose2$ node pairs in an $|N|$-node mesh network, it should be placed at a node with $C_{bw}^{max}$.

\begin{figure}[t]
	\centering
	\includegraphics[width=0.9\linewidth]{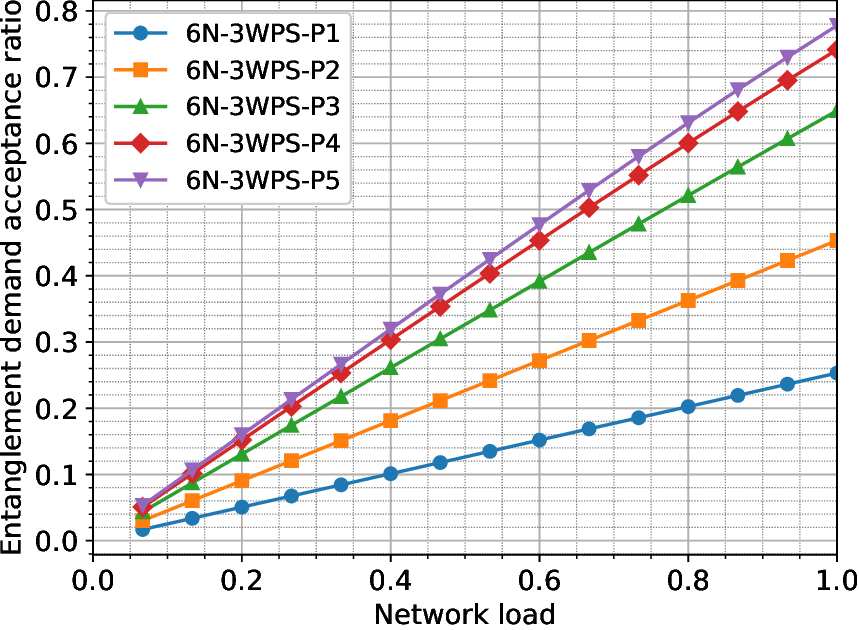}
	\caption{EDAR versus network load for different 3WPS placements represented in Fig. 12 for 6-PMESH network.}
	\label{fig:EDAR_place_3WPS_6PMESH}
\end{figure}

\begin{figure}[t]
	\centering
	\includegraphics[width=0.9\linewidth]{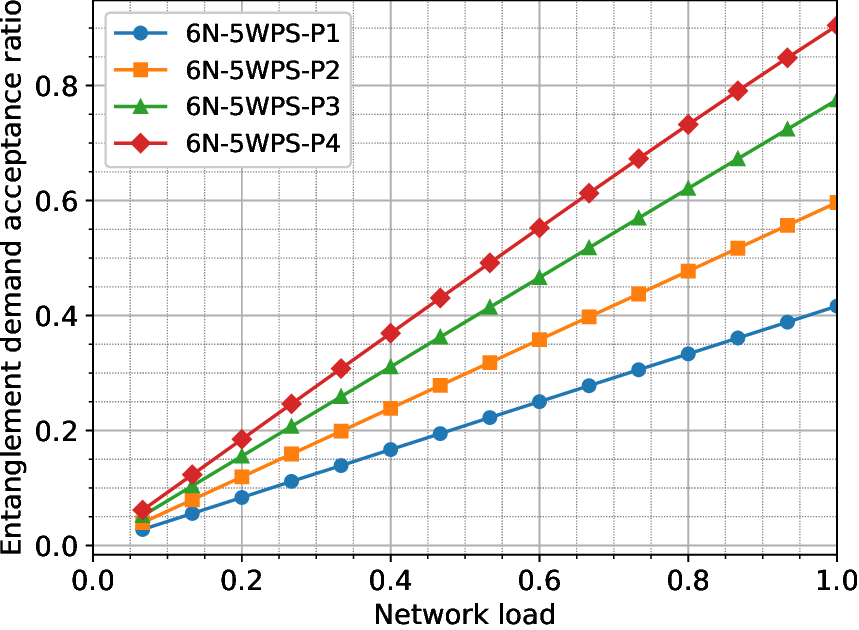}
	\caption{EDAR versus network load for different 5WPS placements represented in Fig. 13 for 6-PMESH network.}
	\label{fig:EDAR_place_5WPS_6PMESH}
\end{figure}

Similar improvements have been observed in EDAR in case of different placements of 3WPSs and 5WPSs in 6-PMESH network topology, as shown in Figs. \ref{fig:EDAR_place_3WPS_6PMESH}-\ref{fig:EDAR_place_5WPS_6PMESH}. In Fig. \ref{fig:EDAR_place_3WPS_6PMESH}, $4.92\%$, $19.79\%$, $71.64\%$, and $207.38\%$ higher EDAR has been achieved for `6N-3WPS-P5' placement as compared to `6N-3WPS-P4', `6N-3WPS-P3', `6N-3WPS-P2', and `6N-3WPS-P1' placements, respectively. In Fig. \ref{fig:EDAR_place_5WPS_6PMESH},  $16.67\%$, $51.61\%$, and $117.22\%$ higher EDAR has been achieved for `6N-5WPS-P4' placement as compared to `6N-5WPS-P3', `6N-5WPS-P2', and `6N-5WPS-P1' placements, respectively. These observations reaffirm that AP distribution is suitable for flexible EPPS placement where, in case of centralized placement of wider-band EPPSs, it can provision most of the entanglement demands based on BP distribution, and in case of diverse placement of narrower-band EPPSs it can provision most of the entanglement demands based on OP.





\begin{figure*}[!t]
\centering
\vspace{-0.2em}
\begin{subfigure}[b]{0.265\textwidth}
  \centering
  \includegraphics[trim=0cm 0cm 0cm 0cm, clip, width=\textwidth]{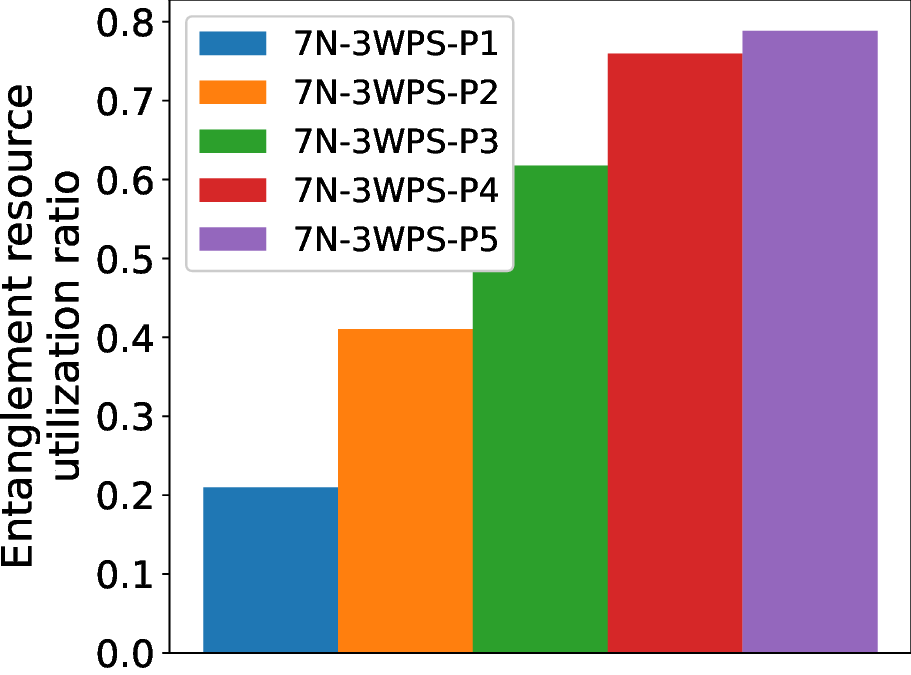}
  \vspace{-1.5em}
  \caption{}
\end{subfigure}\hfill
\begin{subfigure}[b]{0.24\textwidth}
  \centering
  \includegraphics[trim=1.5cm 0cm 0cm 0cm, clip, width=\textwidth]{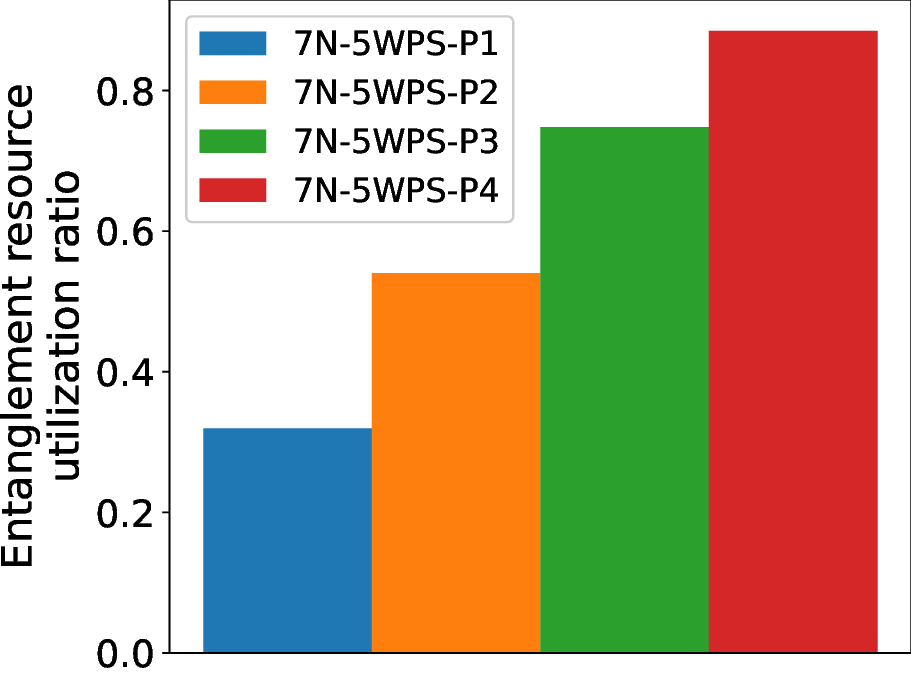}
  \vspace{-1.5em}
  \caption{}
\end{subfigure}\hfill
\begin{subfigure}[b]{0.24\textwidth}
  \centering
  \includegraphics[trim=1.5cm 0cm 0cm 0cm, clip, width=\textwidth]{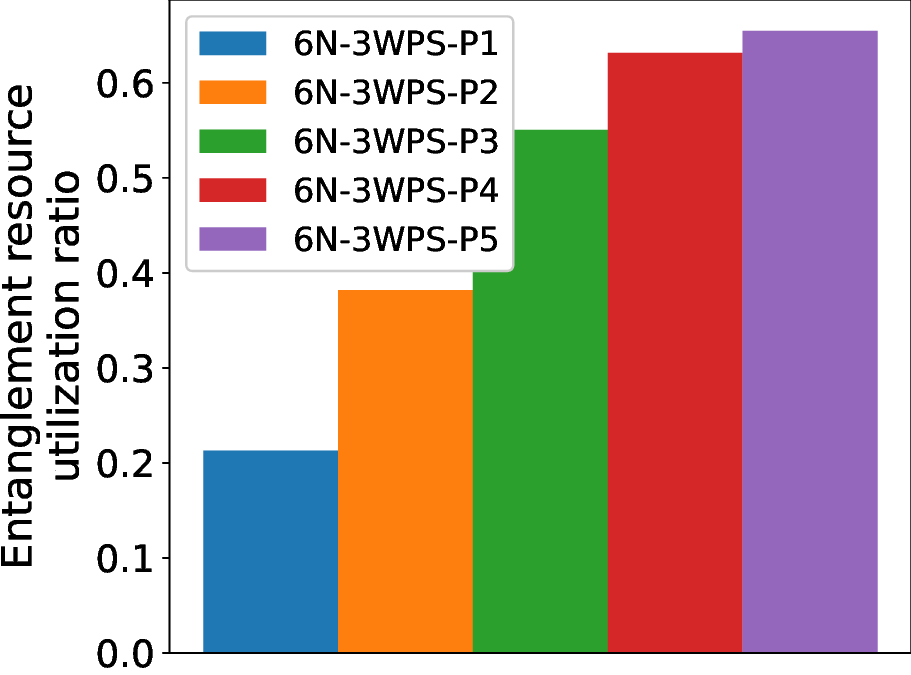}
  \vspace{-1.5em}
  \caption{}
\end{subfigure}\hfill
\begin{subfigure}[b]{0.24\textwidth}
  \centering
  \includegraphics[trim=1.5cm 0cm 0cm 0cm, clip, width=\textwidth]{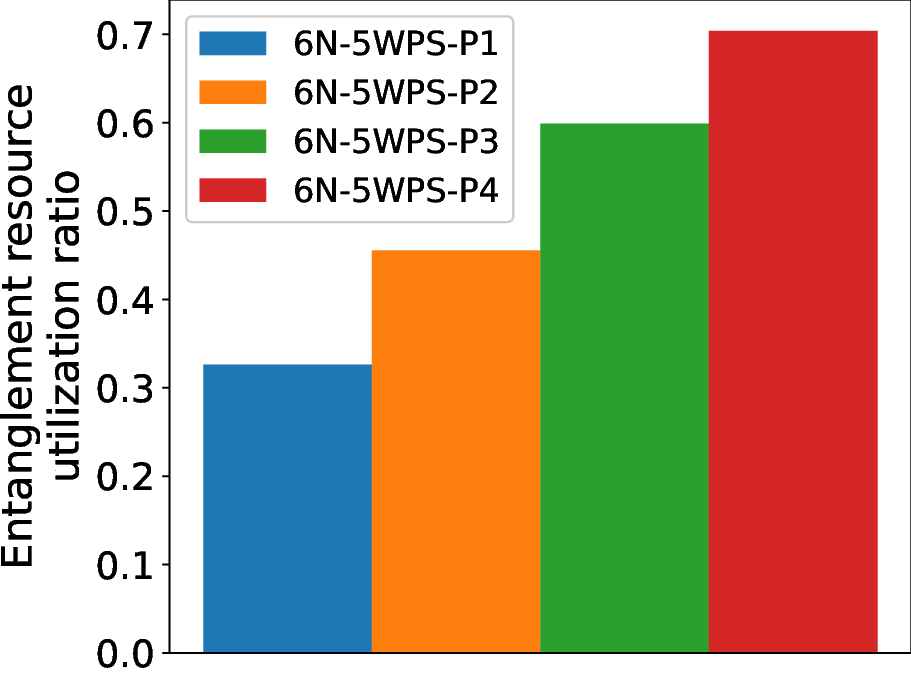}
  \vspace{-1.5em}
  \caption{}
\end{subfigure}
\vspace{-0.3em}
\caption{Entanglement resource utilization ratio (ERUR) at maximum network load for different EPPS placements: (a) 3WPS placements in 7-PMESH network represented in Fig. 10, (b) 5WPS placements in 7-PMESH network represented in Fig. 11, (c) 3WPS placements in 6-PMESH network represented in Fig. 12, and (d) 5WPS placements in 6-PMESH network represented in Fig. 13.}
\label{fig:ERUR_place}
\vspace{-1em}
\end{figure*}

For all the pre-placement variations of considered EPPSs shown in Figs. \ref{fig:place3WPS_7}-\ref{fig:place5WPS_6}, we now analyze entanglement resource utilization ratio (ERUR) as the ratio of the number of entangled photon pairs utilized to the number of entangled photon pairs generated by all the EPPSs in each considered case. The ERUR obtained at maximum network load for all the considered EPPS placement variations is shown in Fig. \ref{fig:ERUR_place}. It is observed that in case of placement of all EPPSs at a single node in the network, placement at a node with $C_{bw}^{max}$ enables higher ERUR (see blue and orange bars in Figs. \ref{fig:ERUR_place}(a)$-$(d)). With distributed placement of EPPSs, ERUR increases with increase in $m$ for all the considered EPPS placement scenarios (see green, red, and purple bars in Figs. \ref{fig:ERUR_place}(a)$-$(d)). The percentage improvements in ERUR are: $3.81\%$, $27.63\%$, $92.13\%$, $275.88\%$ higher for `7N-3WPS-P5' as compared to `7N-3WPS-P4', `7N-3WPS-P3', `7N-3WPS-P2', `7N-3WPS-P1', respectively; $18.33\%$, $63.84\%$, $177\%$ higher for `7N-5WPS-P4' as compared to `7N-5WPS-P3', `7N-5WPS-P2', `7N-5WPS-P1', respectively; $3.67\%$, $18.94\%$, $71.48\%$, $207.63\%$ higher for `6N-3WPS-P5' as compared to `6N-3WPS-P4', `6N-3WPS-P3', `6N-3WPS-P2', `6N-3WPS-P1', respectively; and $17.52\%$, $54.5\%$, $115.84\%$ higher for `6N-5WPS-P4' as compared to `6N-5WPS-P3', `6N-5WPS-P2', `6N-5WPS-P1', respectively, as shown in Figs. \ref{fig:ERUR_place}(a)$-$(d). These percentage improvements in ERUR are in good agreement with the corresponding percentage increases in EDAR for each considered case, as observed in Figs. \ref{fig:EDAR_place_3WPS_7PMESH}-\ref{fig:EDAR_place_5WPS_6PMESH}. Thus, distributed EPPS placement (based on dominant set) at nodes with maximum betweenness centrality achieve high ERUR due to lower loss of photon pairs and visibility from source node/s to the remaining node/s in the network that subsequently results in high EDAR.

\section{Conclusions}

We presented a distributed entanglement distribution approach in WDM-based mesh quantum optical networks using multiple EPPSs. We experimentally characterized two EPPSs developed in-house and performed network simulations considering both the EPPSs individually. We analyzed different entanglement distribution schemes, namely, OP distribution that is commonly used in point-to-point studies, and BP distribution that is commonly used in star network studies, and observed that a hybrid scheme of OP and BP, namely, AP is advantageous for entanglement distribution with multiple EPPSs in mesh quantum optical networks. Based on a comprehensive analysis done for several network topologies and EPPS placement variations, it has been observed that distributed placement of multiple EPPSs based on betweenness centrality and dominant sets enables higher entanglement resource utilization and achieves higher entanglement demand acceptance as compared to centralized placement. Thus, we conclude that similar to the application-level advantages of distributivity in quantum networks such as distributed quantum computing and sensing, distributed entanglement distribution using multiple EPPSs is advantageous as compared to centralized entanglement distribution. 


\bibliographystyle{IEEEtran}
\small
\bibliography{references}

\end{document}